\pdfoutput=1
\documentclass[letterpaper,english,11pt,aps,letter,superscriptaddress,nofootinbib,floatfix,pre]{revtex4}
\usepackage[T1]{fontenc}
\usepackage[latin1]{inputenc}
\usepackage{amsmath}
\usepackage{color}
\usepackage{graphicx}
\usepackage{amssymb}

\makeatletter
\usepackage{ae,aecompl}

\renewcommand{\citet}[1]{\cite{#1}}

\makeatother

\usepackage{babel}

\begin{document}

\title{On the Accuracy of Finite-Volume Schemes for Fluctuating Hydrodynamics}

\author{Aleksandar Donev}

\email{aleks.donev@gmail.com}

\affiliation{Lawrence Livermore National Laboratory, P.O.Box 808, Livermore, CA
94551-9900}

\affiliation{Center for Computational Science and Engineering, Lawrence Berkeley
National Laboratory, Berkeley, CA, 94720}

\author{Eric Vanden-Eijnden}

\affiliation{Courant Institute of Mathematical Sciences, New York University,
New York, NY 10012}

\author{Alejandro L. Garcia}

\affiliation{Department of Physics, San Jose State University, San Jose, California,
95192}

\author{John B. Bell}

\affiliation{Center for Computational Science and Engineering, Lawrence Berkeley
National Laboratory, Berkeley, CA, 94720}

\begin{abstract}
This paper describes the development and analysis of finite-volume
methods for the Landau-Lifshitz Navier-Stokes (LLNS) equations and
related stochastic partial differential equations in fluid dynamics.
The LLNS equations incorporate thermal fluctuations into macroscopic
hydrodynamics by the addition of white-noise fluxes whose magnitudes
are set by a fluctuation-dissipation relation. Originally derived
for equilibrium fluctuations, the LLNS equations have also been shown
to be accurate for non-equilibrium systems. Previous studies of numerical
methods for the LLNS equations focused primarily on measuring variances
and correlations computed at equilibrium and for selected non-equilibrium
flows. In this paper, we introduce a more systematic approach based
on studying discrete equilibrium structure factors for a broad class
of explicit linear finite-volume schemes. This new approach provides
a better characterization of the accuracy of a spatio-temporal discretization
as a function of wavenumber and frequency, allowing us to distinguish
between behavior at long wavelengths, where accuracy is a prime concern,
and short wavelengths, where stability concerns are of greater importance.
We use this analysis to develop a specialized third-order Runge Kutta
scheme that minimizes the temporal integration error in the discrete
structure factor at long wavelengths for the one-dimensional linearized
LLNS equations. Together with a novel method for discretizing the
stochastic stress tensor in dimension larger than one, our improved
temporal integrator yields a scheme for the three-dimensional equations
that satisfies a discrete fluctuation-dissipation balance for small
time steps and is also sufficiently accurate even for time steps close
to the stability limit.
\end{abstract}
\maketitle
\newcommand{\Cross}[1]{\left|\boldsymbol{#1}\right|_{\times}}
\newcommand{\CrossL}[1]{\left|\boldsymbol{#1}\right|_{\times}^{L}}
\newcommand{\CrossR}[1]{\left|\boldsymbol{#1}\right|_{\times}^{R}}
\newcommand{\CrossS}[1]{\left|\boldsymbol{#1}\right|_{\boxtimes}}

\newcommand{\V}[1]{\boldsymbol{#1}}
\newcommand{\M}[1]{\boldsymbol{#1}}
\newcommand{\D}[1]{\Delta#1}

\newcommand{\Set}[1]{\mathbb{#1}}

\newcommand{\ki}{k}
\newcommand{\wi}{\omega}

\newcommand{\grad}{\boldsymbol{\nabla}}
\newcommand{\eij}{\left\{  i,j\right\}  }

\newcommand{\Wi}{\mbox{Wi}}
\newcommand{\modified}[1]{\textcolor{red}{#1}}

\section{Introduction}

Recently the fluid dynamics community has considered increasingly
complex physical, chemical, and biological phenomena at the microscopic
scale, including systems for which significant interactions occur
across multiple scales. At a molecular scale, fluids are not deterministic;
the state of the fluid is constantly changing and stochastic, even
at thermodynamic equilibrium. As simulations of fluids push toward
the microscale, these random thermal fluctuations play an increasingly
important role in describing the state of the fluid, especially when
investigating systems where the microscopic fluctuations drive a macroscopic
phenomenon such as the evolution of instabilities, or where the thermal
fluctuations drive the motion of suspended microscopic objects in
complex fluids. Some examples in which spontaneous fluctuations can
significantly affect the dynamics include the breakup of droplets
in jets \citet{Moseler:00,Eggers:02,Kang:07}, Brownian molecular
motors \citet{Astumian:02,Oster:02,Broeck:04,Meurs:04}, Rayleigh-Bernard
convection (both single species \citet{Wu:95} and mixtures \citet{Quentin:95},
Kolmogorov flows \citet{Bena:00,Bena:99,Mansour:99}, Rayleigh-Taylor
mixing \citet{Kadau:04,FluidMixing_DSMC}, combustion and explosive
detonation \citet{Nowakowski:03,Lemarchand:04}, and reaction fronts
\citet{Moro:04}.

Numerical schemes based on a particle representation of a fluid (e.g.,
molecular dynamics, direct simulation Monte Carlo \citet{DSMCReview_Garcia})
inherently include spontaneous fluctuations due to the irregular dynamics
of the particles. However, by far the most common numerical schemes
in computational fluid dynamics are based on solving partial differential
equations. To incorporate thermal fluctuations into macroscopic hydrodynamics,
Landau and Lifshitz introduced an extended form of the compressible
Navier-Stokes equations obtained by adding white-noise stochastic
flux terms to the standard deterministic equations. While they were
originally developed for equilibrium fluctuations, specifically the
Rayleigh and Brillouin spectral lines in light scattering, the validity
of the Landau-Lifshitz Navier-Stokes (LLNS) equations for non-equilibrium
systems has been assessed \citet{LLNS_Espanol} and verified in molecular
simulations \citet{Garcia:91,Mansour:87,Mareschal:92}. The LLNS system
is one of the more complex examples in a broad family of PDEs with
stochastic fluxes. Many members of this family arise from the LLNS
equations in a variety of approximations (e.g., stochastic heat equation)
while others are stochastic variants of well-known PDEs, such as the
stochastic Burger's equation \citet{Bell:06}, which can be derived
from the continuum limit of an asymmetric excluded random walk.

Several numerical approaches for fluctuating hydrodynamics have been
proposed. The earliest work by Garcia \emph{et al}. \citet{Garcia:87}
developed a simple scheme for the stochastic heat equation and the
linearized one-dimensional LLNS equations. Ladd \emph{et al.} have
included stress fluctuations in (isothermal) Lattice-Boltzmann methods
for some time \citet{Ladd:93}, and recently a better theoretical
foundation has been established \citet{LB_ThermalFluctuations,LB_StatMech}.
Moseler and Landman \citet{Moseler:00} included the stochastic stress
tensor of the LLNS equations in the lubrication equations and obtain
good agreement with their molecular dynamics simulation in modeling
the breakup of nano-jets. Sharma and Patankar \citet{Sharma:04} developed
a fluid-structure coupling between a fluctuating incompressible solver
and suspended Brownian particles. Coveney, De Fabritiis, Delgado-Buscalioni
and co-workers have also used the isothermal LLNS equations in a hybrid
scheme, coupling a continuum fluctuating solver to a Molecular Dynamics
simulation of a liquid \citet{FluctuatingHydro_Coveney,FluctuatingHydroMD_Coveney,FluctuatingHydroHybrid_MD}.
Atzberger and collaborators \citet{AtzbergerETAL:2007} have developed
a version of the immersed boundary method that includes fluctuations
in a pseudo-spectral method for the incompressible Navier-Stokes equations.
Voulgarakis and Chu \citet{StagerredFluctHydro} developed a staggered
scheme for the isothermal LLNS equations as part of a multiscale method
for biological applications, and a similar staggered scheme was also
described in Ref. \citet{Delgado:08}.

Recently, Bell \emph{et al.} \citet{Bell:07} introduced a centered
scheme for the LLNS equations based on interpolation schemes designed
to preserve fluctuations combined with a third-order Runge-Kutta (RK3)
temporal integrator. In that work, the principal diagnostic used for
evaluation of the numerical method was the accuracy of the local (cell)
variance and spatial (cell-to-cell) correlation structure for equilibrium
and selected non-equilibrium scenarios (e.g., constant temperature
gradient). The metric established by those types of tests is, in some
sense, simultaneously too crude and too demanding. It is too crude
in the sense that it provides only limited information from detailed
simulations that cannot be directly linked to specific properties
of the scheme. On the other hand, such criteria are too demanding
in the sense that they place requirements on the discretization integrated
over all wavelengths, requiring that the method perform well at high
wavenumbers where a deterministic PDE solver performs poorly. Furthermore,
although Bell \emph{et al.} \citet{Bell:07} demonstrate that RK3
is an effective algorithm, compared with other explicit schemes for
the compressible Navier-Stokes equations, the general development
of schemes for the LLNS equations has been mostly trial-and-error.

Here, our goal is to establish a more rational basis for the analysis
and development of explicit finite volume scheme for SPDEs with a
stochastic flux. The approach is based on analysis of the structure
factor (equilibrium fluctuation spectrum) of the discrete system.
The structure factor is, in essence, the stationary spatio-temporal
correlations of hydrodynamic fluctuations as a function of spatial
wavenumber and temporal frequency; the static structure factor is
the integral over frequency (i.e., the spatial spectrum). By analyzing
the structure factor for a numerical scheme, we are able to develop
notions of accuracy for a given discretization at long wavelengths.
Furthermore, in many cases the theoretical analysis for the structure
factor is tractable (with the aid of symbolic manipulators) allowing
us to determine optimal coefficients for a given numerical scheme.
We perform this optimization as a two-step procedure. First, a spatial
discretization is developed that satisfies a discrete form of the
fluctuation-dissipation balance condition. Then, a stable temporal
integrator is proposed and the covariances of the random numbers are
chosen so as to maximize the order of temporal accuracy of the small-wavenumber
static structure factor. We focus primarily on explicit schemes for
solving the LLNS equations because even at the scales where thermal
fluctuations are important, the limitation on time step imposed by
stability is primarily due to the hyperbolic terms. That is, when
the cell size is comparable to the length scale for molecular transport
(e.g., mean free path in a dilute gas) the time step for these compressible
hydrodynamic equations is limited by the acoustic CFL condition. At
even smaller length scales the viscous terms further limit the time
step yet the validity of a continuum representation for the fluid
starts to break down at those atomic scales.

The paper is divided into roughly two parts: The first half (sections
\ref{sec:Landau-Lifshitz-Navier-Stokes-Equations}-\ref{SectionLinearGeneral})
defines notation, develops the formalism, and derives the expressions
for analyzing a general class of linear stochastic PDEs from the LLNS
family of equations. The main result in the first half, how to evaluate
the structure factor for a numerical scheme, appears in section \ref{sub:Analysis-of-Linear-Explicit-Methods}.
The second half applies this analysis to systems of increasing complexity,
starting with the stochastic heat equation (section \ref{Section_S_k_diffusion}),
followed by the LLNS system in one dimension (section \ref{sec:Section-LLNS-Equations-1D})
and three dimensions (section \ref{sec:Higher-Dimensions}). The paper
closes with a summary and concluding remarks.

\section{Landau-Lifshitz Navier-Stokes Equations\label{sec:Landau-Lifshitz-Navier-Stokes-Equations}}

We consider the accuracy of explicit finite-volume methods for solving
the Landau-Lifshitz Navier-Stokes (LLNS) system of stochastic partial
differential equations (SPDEs) in $d$ dimensions, given in conservative
form by\begin{equation}
\partial_{t}\V{U}=-\grad\cdot\left[\V{F}(\V{U})-\V{\mathcal{Z}}(\V{U},\V{r},t)\right],\label{LLNS_general}\end{equation}
where $\V{U}\left(\V{r},t\right)=\left[\begin{array}{ccc}
\rho, & \V{j}, & e\end{array}\right]^{T}$ is a vector of \emph{conserved variables} that are a function of
the spatial position $\V{r}$ and time $t$. The conserved variables
are the densities of mass $\rho$, momentum $\V{j}=\rho\V{v}$, and
energy $e=\epsilon(\rho,T)+\frac{1}{2}\rho v^{2}$, expressed in terms
of the \emph{primitive variables}, mass density $\rho$, velocity
$\V{v}$ and temperature $T$; here $\epsilon$ is the internal energy
density. The deterministic flux is taken from the traditional compressible
Navier-Stokes-Fourier equations and can be split into \emph{hyperbolic}
and \emph{diffusive fluxes}:\[
\V{F}(\V{U})=\V{F}_{H}(\V{U})+\V{F}_{D}(\V{U}),\]
where \[
\V{F}_{H}=\left[\begin{array}{c}
\rho\V{v}\\
\rho\V{v}\V{v}^{T}+P\M{I}\\
(e+P)\V{v}\end{array}\right]\mbox{ and }\V{F}_{D}=-\left[\begin{array}{c}
\V{0}\\
\M{\sigma}\\
\M{\sigma}\cdot\V{v}+\M{\xi}\end{array}\right],\]
$P=P(\rho,T)$ is the pressure, the viscous stress tensor is $\M{\sigma}=2\eta\left[\frac{1}{2}(\grad\V{v}+\grad\V{v}^{T})-\frac{\left(\grad\cdot\V{v}\right)}{d}\M{I}\right]$
for $d\geq2$ (we have assumed zero bulk viscosity) and $\M{\sigma}=\eta v_{x}$
for $d=1$, and the heat flux is $\M{\xi}=\mu\grad T$. We denote
the adjoint (conjugate transpose) of a matrix or linear operator $\M{M}$
with $\M{M}^{\star}=\overline{\M{M}}^{T}$. As postulated by Landau-Lifshitz
\citet{Landau:Fluid,LLNS_Espanol}, the \emph{stochastic flux} \[
\V{\mathcal{Z}}=\left[\begin{array}{c}
\V{0}\\
\M{\Sigma}\\
\M{\Sigma}\cdot\V{v}+\M{\Xi}\end{array}\right]\]
is composed of the stochastic stress tensor $\M{\Sigma}$ and stochastic
heat flux vector $\M{\Xi}$, assumed to be mutually uncorrelated random
Gaussian fields with a covariance\begin{align}
\left\langle \M{\Sigma}(\V{r},t)\M{\Sigma}^{\star}(\V{r}^{\prime},t^{\prime})\right\rangle = & \M{C}_{\M{\Sigma}}\delta(t-t^{\prime})\delta(\V{r}-\V{r}^{\prime})\mbox{, where }C_{ij,kl}^{(\M{\Sigma})}=2\bar{\eta}k_{B}\overline{T}\left(\delta_{ik}\delta_{jl}+\delta_{il}\delta_{jk}-\frac{2}{d_{f}}\delta_{ij}\delta_{kl}\right)\nonumber \\
\left\langle \M{\Xi}(\V{r},t)\M{\Xi}^{\star}(\V{r}^{\prime},t^{\prime})\right\rangle = & \M{C}_{\M{\Xi}}\delta(t-t^{\prime})\delta(\V{r}-\V{r}^{\prime})\mbox{, where }C_{i,j}^{(\M{\Xi})}=2\bar{\mu}k_{B}\overline{T}^{2}\delta_{ij}\label{stoch_flux_covariance}\end{align}

In the LLNS system, the \emph{hyperbolic} or \emph{advective} fluxes
are responsible for transporting the conserved quantities at the speed
of sound or fluid velocity, without dissipation. On the other hand,
the \emph{diffusive} or \emph{dissipative} fluxes are the ones responsible
for damping the thermal fluctuations generated by the \emph{stochastic
}or \emph{fluctuating }fluxes. At equilibrium a steady state is reached
in which a \emph{fluctuation-dissipation balance} condition is satisfied.

In the original formulation, Landau and Lifshitz only considered adding
stochastic fluxes to the linearized Navier-Stokes equations, which
leads to a well-defined system of SPDEs whose equilibrium solutions
are random Gaussian fields. Derivations of the equations of fluctuating
hydrodynamics through careful asymptotic expansions of the underlying
microscopic (particle) dynamics give equations for the Gaussian fluctuations
around the solution to the usual deterministic Navier-Stokes equations
\citet{LebowitzHydroReview}, in the spirit of the Central Limit Theorem.
Therefore, numerical solutions should, in principle, consist of two
steps: First solving the nonlinear deterministic equations for the
\emph{mean} solution, and then solving the linearized equations for
the \emph{fluctuations} around the mean. If the fluctuations are small
perturbations, it makes sense numerically to try to combine these
two steps into one and simply consider non-linear equations with added
thermal fluctuations. There is also hope that this might capture effects
not captured in the two-system approach, such as the effect of fluctuations
on the very long-time dynamics of the mean (e.g., shock drift \citet{Bell:07})
or hydrodynamic instabilities \citet{Wu:95,Moseler:00,FluidMixing_DSMC}.

The linearized equations of fluctuating hydrodynamics can be given
a well-defined interpretation with the use of generalized functions
or distributions \citet{DaPratoBook}. However, the non-linear fluctuating
hydrodynamic equations (\ref{LLNS_general}) must be treated with
some care since they have not been derived from first-principles \citet{LLNS_Espanol}
and are in fact mathematically ill-defined due to the high irregularity
of white-noise fluctuating stresses \citet{GardinerBook}. More specifically,
because the solution of these equations is itself a distribution the
interpretation of the nonlinear terms requires giving a precise meaning
to products of distributions, which cannot be defined in general and
requires introducing some sort of regularization. Although written
formally as an SPDE, the LLNS equations are usually interpreted in
a finite volume context, where the issues of regularity, at first
sight, disappear. However, in finite volume form the level of fluctuations
becomes increasingly large as the volume shrinks and the non-linear
terms diverge leading to an {}``ultraviolet catastrophe'' of the
kind familiar in other fields of physics. Furthermore, because the
noise terms are Gaussian, it is possible for rare events to push the
system to states that are not thermodynamically valid such as negative
$T$ or $\rho$. For that reason, we will focus on the linearized
LLNS equations, which can be given a well-defined interpretation.
Since the fluctuations are expected to be a small perturbation of
the deterministic solution, the nonlinear equations should behave
similarly to the linearized equations anyway, at least near equilibrium
for sufficiently large cells.

To simplify the exposition we assume the fluid to be a mono-atomic
ideal gas; the generalization of the results for an arbitrary fluid
is tedious but straightforward. For an ideal gas the equation of state
may be written as $P=\rho\left(k_{B}T/m\right)=\rho c^{2}$, where
$c$ is the isothermal speed of sound. The internal energy density
is $\epsilon=\rho c_{v}T$, where $c_{v}$ is the heat capacity at
constant volume, which may be written as $c_{v}=d_{f}k_{B}/2m$ where
$d_{f}$ is the number of degrees of freedom of the molecules (for
monoatomic gases there are $d_{f}=d$ translational degrees of freedom),
and $c_{p}=(1+2/d_{f})c_{v}$ is the heat capacity at constant pressure.
For analytical calculations it is convenient to convert the LLNS system
from conserved variables to primitive variables, since the primitive
variables are uncorrelated at equilibrium and the equations (\ref{LLNS_general})
simplify considerably,\begin{align}
D_{t}\rho= & \grad\ \cdot\left(\rho\V{v}\right)\nonumber \\
\rho\left(D_{t}\V{v}\right)= & -\grad P+\grad\cdot\left(\M{\sigma}+\M{\Sigma}\right)\nonumber \\
\rho c_{p}\left(D_{t}T\right)= & D_{t}P+\grad\cdot\left(\M{\xi}+\M{\Xi}\right)+\left(\M{\sigma}+\M{\Sigma}\right):\grad\V{v},\label{LLNS_primitive}\end{align}
where $D_{t}\square=\partial_{t}\square+\V{v}\cdot\grad\left(\square\right)$
denotes the familiar advective derivative. Note that in the fully
non-linear numerical implementation, however, we continue to use the
conserved variables to ensure that the physical conservation laws
are strictly obeyed.

Linearizing (\ref{LLNS_primitive}) around a reference uniform equilibrium
state $\rho=\rho_{0}+\delta\rho$, $\V{v}=\V{v}_{0}+\delta\V{v}$,
$T=T_{0}+\delta T$, and dropping the deltas for notational simplicity,\[
\V{U}=\left[\begin{array}{c}
\delta\rho\\
\delta\V{v}\\
\delta T\end{array}\right]\rightarrow\left[\begin{array}{c}
\rho\\
\V{v}\\
T\end{array}\right],\]
we obtain the linearized LLNS system for the equilibrium thermal fluctuations,\begin{equation}
\partial_{t}\V{U}=-\grad\cdot\left[\M{F}\V{U}-\V{\mathcal{Z}}\right]=-\grad\cdot\left[\M{F}_{H}\V{U}+\M{F}_{D}\grad\V{U}-\V{\mathcal{Z}}\right],\label{LLNS_linear_ideal}\end{equation}
where\[
\M{F}_{H}\V{U}=\left[\begin{array}{c}
\rho_{0}\V{v}+\rho\V{v}_{0}\\
\left(c_{0}^{2}\rho_{0}^{-1}\rho+c_{0}^{2}T_{0}^{-1}T\right)\M{I}+\V{v}_{0}\V{v}^{T}\\
c_{0}^{2}c_{v}^{-1}\V{v}+T\V{v}_{0}\end{array}\right]\mbox{ and }\M{F}_{D}\grad\V{U}=\left[\begin{array}{c}
0\\
\rho_{0}^{-1}\eta_{0}\overline{\grad}\V{v}\\
\rho_{0}^{-1}c_{v}^{-1}\mu_{0}\grad T\end{array}\right],\]
and $\overline{\grad}$ denotes a symmetrized traceless gradient,
$\overline{\grad}\V{v}=\frac{1}{2}(\grad\V{v}+\grad\V{v}^{T})-\frac{\left(\grad\cdot\V{v}\right)}{d}\M{I}$.
Here $\V{\mathcal{Z}}(\V{r},t)$ is a random Gaussian field with a
covariance\[
\left\langle \V{\mathcal{Z}}(\V{r},t)\V{\mathcal{Z}}^{\star}(\V{r}^{\prime},t^{\prime})\right\rangle =\M{C}_{\V{Z}}\delta(t-t^{\prime})\delta(\V{r}-\V{r}^{\prime}),\]
where the covariance matrix is block diagonal,\[
\M{C}_{\V{Z}}=\left[\begin{array}{ccc}
0 & \V{0} & \V{0}\\
\V{0} & \rho_{0}^{-2}\M{C}_{\M{\Sigma}} & \M{0}\\
\V{0} & \M{0} & \rho_{0}^{-2}c_{v}^{-2}\M{C}_{\M{\Xi}}\end{array}\right],\]
and $\M{C}_{\M{\Sigma}}$ and $\M{C}_{\M{\Xi}}$ are given in Eq.
(\ref{stoch_flux_covariance}). Equation (\ref{LLNS_linear_ideal})
is a system of linear SPDEs with additive noise that can be analyzed
within a general framework, as we develop next. We note that the stochastic
{}``forcing'' in (\ref{LLNS_linear_ideal}) is essentially a divergence
of white noise, modeling conservative \emph{intrinsic} (thermal) fluctuations
\citet{LebowitzHydroReview}, rather than the more common \emph{external}
fluctuations modeled through white noise forcing \citet{StochHeatEq_Weak}.

The next two sections develop the tools for analyzing explicit finite
volume schemes for linearized SPDEs, such as the LLNS system, specifically
how to predict the equilibrium spectrum of the fluctuations (i.e.,
structure factor) from the spatial and temporal discretization used
by the numerical algorithm. These analysis tools are demonstrated
for simple examples in Section \ref{Section_S_k_diffusion} and applied
to the LLNS system in Sections \ref{sec:Section-LLNS-Equations-1D}
and \ref{sec:Higher-Dimensions}.

\section{\label{sec:Explicit-Methods}Explicit Methods for Linear Stochastic
Partial Differential Equations}

In this section, we develop an approach for analyzing the behavior
of explicit discretizations for a broad class of SPDEs, motivated
by the linearized form of the LLNS equations. In particular, we consider
a general linear SPDE for the stochastic field $\V{\mathcal{U}}(\V{r},t)\equiv\V{\mathcal{U}}(t)$
of the form\begin{equation}
d\V{\mathcal{U}}(t)=\M{\mathcal{L}}\V{\mathcal{U}}(t)dt+\M{\mathcal{K}}d\V{\mathcal{B}}(t),\label{U_t_linear_general}\end{equation}
with periodic boundary conditions on the torus $\V{r}\in\mathcal{V}=\left[0,H\right]^{d}$,
where $\M{\mathcal{L}}$ (the \emph{generator}) and $\M{\mathcal{K}}$
(the \emph{filter}) are time-independent linear operators, and $\V{\mathcal{B}}$
is a cylindrical Wiener process (Brownian sheet), and the initial
condition at $t=0$ is $\V{\mathcal{U}}_{0}$. As common in the physics
literature, we will abuse notation and often write\[
\partial_{t}\V{\mathcal{U}}=\M{\mathcal{L}}\V{\mathcal{U}}+\M{\mathcal{K}}\V{\mathcal{W}},\]
where $\V{\mathcal{W}}=d\V{\mathcal{B}}(t)/dt$ is spatio-temporal
white noise, i.e., a random Gaussian field with zero mean and covariance
\begin{equation}
\left\langle \V{\mathcal{W}}(\V{r},t)\V{\mathcal{W}}^{\star}(\V{r}^{\prime},t^{\prime})\right\rangle =\delta(t-t^{\prime})\delta(\V{r}-\V{r}^{\prime}).\label{White_Noise_Covariance}\end{equation}

The so-called mild solution \citet{DaPratoBook} of (\ref{U_t_linear_general})
is a generalized process \begin{equation}
\V{\mathcal{U}}(t)=e^{t\M{\mathcal{L}}}\V{\mathcal{U}}_{0}+\int_{0}^{t}e^{(t-s)\M{\mathcal{L}}}\M{\mathcal{K}}d\V{\mathcal{B}}(s),\label{U_linear_formal_sol}\end{equation}
where the integral denotes a stochastic convolution. If the operator
$\M{\mathcal{L}}$ is dissipative, that is, $e^{t\M{\mathcal{L}}}\V{\mathcal{U}}_{0}\underset{t\rightarrow\infty}{\rightarrow}\V{0}$
for all $\V{\mathcal{U}}_{0}$, then at long times $t^{\prime}$ the
solution to (\ref{U_t_linear_general}) is a Gaussian process with
mean zero and covariance\begin{equation}
\M{C}_{\V{\mathcal{U}}}(t)=\left\langle \V{\mathcal{U}}(t^{\prime})\V{\mathcal{U}}^{\star}(t^{\prime}+t)\right\rangle =\int_{-\infty}^{0}e^{-s\M{\mathcal{L}}}\M{\mathcal{K}}\M{\mathcal{K}}^{\star}e^{(t-s)\M{\mathcal{L}}^{\star}}ds,\quad t\ge0.\label{C_U_general_continuum}\end{equation}
This means that (\ref{U_t_linear_general}) has a unique invariant
measure (equilibrium or stationary distribution) that is Gaussian
with mean zero and covariance given in Eq. (\ref{C_U_general_continuum}).

In general, the field $\V{\mathcal{U}}(\V{r},t)$ is only a generalized
function of the spatial coordinate $\V{r}$ and cannot be evaluated
pointwise. For the cases we will consider here, specifically, translationally-invariant
problems where $\M{\mathcal{L}}$ and $\M{\mathcal{K}}$ are differential
operators, this difficulty can be avoided by transforming (\ref{U_t_linear_general})
to Fourier space via the Fourier series transform\begin{align}
\V{\mathcal{U}}(\V{r},t)= & \sum_{\V{k}\in\widehat{\mathcal{V}}}e^{i\V{k}\cdot\V{r}}\widehat{\V{\mathcal{U}}}(\V{k},t)\label{Cont_FT_inverse}\\
\widehat{\V{\mathcal{U}}}(\V{k},t)=\frac{1}{V} & \int_{\V{r}\in\mathcal{V}}e^{-i\V{k}\cdot\V{r}}\V{\mathcal{U}}(\V{r},t)d\V{r},\label{Cont_FT_forward}\end{align}
where $V=\left|\mathcal{V}\right|=H^{d}$ is the volume of the system,
and each \emph{wavevector} $\V{k}\equiv\V{k}(\V{\kappa})$ is expressed
in terms of the integer \emph{wave index} $\V{\kappa}\in\Set{Z}^{d}$,
giving the set of discrete wavevectors \[
\widehat{\mathcal{V}}=\left\{ \V{k}=2\pi\V{\kappa}/H\quad\left|\quad\right.\V{\kappa}\in\Set{Z}^{d}\right\} .\]
In Fourier space, the SPDE (\ref{U_t_linear_general}) becomes an
infinite system of uncoupled stochastic ordinary differential equations
(SODEs),\begin{equation}
d\widehat{\V{\mathcal{U}}}(t)=\widehat{\M{\mathcal{L}}}\widehat{\V{\mathcal{U}}}(t)dt+\widehat{\M{\mathcal{K}}}d\widehat{\V{\mathcal{B}}}(t),\label{U_hat_t_linear_general}\end{equation}
one SODE for each $\V{k}\in\widehat{\mathcal{V}}$ . The invariant
distribution of (\ref{U_hat_t_linear_general}) is a zero-mean Gaussian
random process, characterized fully by the covariance obtained from
the spatial Fourier transform of (\ref{C_U_general_continuum}),\begin{equation}
\M{\mathcal{S}}(\V{k},t)=V\left\langle \widehat{\V{\mathcal{U}}}(\V{k},t^{\prime})\widehat{\V{\mathcal{U}}}^{\star}(\V{k},t^{\prime}+t)\right\rangle =\frac{1}{2\pi}\int_{-\infty}^{\infty}e^{i\omega t}\M{\mathcal{S}}(\V{k},\omega)d\omega,\label{S_k_t_cont_def}\end{equation}
where the \emph{dynamic structure factor} (space-time spectrum) is
\begin{equation}
\M{\mathcal{S}}(\V{k},\omega)=V\left\langle \widehat{\V{\mathcal{U}}}(\V{k},\omega)\widehat{\V{\mathcal{U}}}^{\star}(\V{k},\omega)\right\rangle =\left(\widehat{\M{\mathcal{L}}}-i\omega\right)^{-1}\left(\widehat{\M{\mathcal{K}}}\widehat{\M{\mathcal{K}}}^{\star}\right)\left(\widehat{\M{\mathcal{L}}}^{\star}+i\omega\right)^{-1},\label{S_U_dyn_continuum}\end{equation}
which follows directly from the space-time $(\V{k},\omega)$ Fourier
transform of the SPDE (\ref{U_t_linear_general}). By integrating
the dynamic spectrum over all frequencies $\omega$, one gets the
\emph{static structure factor} \begin{equation}
\M{\mathcal{S}}(\V{k})=\M{\mathcal{S}}(\V{k},t=0)=\frac{1}{2\pi}\int_{-\infty}^{\infty}\M{\mathcal{S}}(\V{k},\omega)d\omega,\label{S_U_static_continuum}\end{equation}
which is the spatial spectrum of an equilibrium snapshot of the fluctuating
field and is the Fourier equivalent of $\M{C}_{\V{\mathcal{U}}}(t=0)$.
Note that the static structure factor of spatial white noise (a snapshot
of $\V{\mathcal{W}}$) is unity independent of the wavevector, $\M{\mathcal{S}}_{\V{\mathcal{W}}}(\V{k})=V\left\langle \V{\mathcal{W}}(\V{k},t)\V{\mathcal{W}}^{\star}(\V{k},t)\right\rangle =\M{I}$.

\subsection{Discretization}

For the types of equations we will consider in this paper, the invariant
measure is spatially white, specifically, $\M{S}(\V{k})$ is diagonal
and independent of $\V{k}$. The associated fluctuating field $\V{\mathcal{U}}$
cannot be evaluated pointwise, therefore, it is more natural to use
\emph{finite-volume} cell averages, denoted here by $\V{U}$. In the
deterministic setting, for uniform periodic grids there is no important
difference between finite-volume and finite-difference methods. Our
general approach can likely be extended also to analysis of stochastic
finite-element discretizations, however, such methods have yet to
be developed for the LLNS equations and here we focus on finite-volume
methods. For notational simplicity, we will discuss problems in one
spatial dimension ($d=1$), with (mostly) obvious generalizations
to higher dimensions.

Space is discretized into $N_{c}$ identical cells of length $\D{x}=H/N_{c}$,
and the value $\V{U}_{j}$ stored in cell $1\leq j\leq N_{c}$ is
the average of the corresponding variable over the cell\begin{equation}
\V{U}_{j}(t)=\frac{1}{\D{x}}\int_{(j-1)\D{x}}^{j\D{x}}\V{\mathcal{U}}(x,t)dx.\label{U_j_finite_volume}\end{equation}
Time is discretized with a time step $\D{t}$, approximating cell
averages of $\V{\mathcal{U}}(x,t)$ pointwise in time with $\V{U}^{n}=\left\{ \V{U}_{1}^{n},...,\V{U}_{N_{c}}^{n}\right\} $,\[
\V{U}_{j}^{n}\approx\V{U}_{j}(n\D{t}),\]
where $n\geq0$ enumerates the time steps. The white noise $\V{\mathcal{W}}(x,t)$
cannot be evaluated pointwise in either space or time and is discretized
using a spatio-temporal average\begin{equation}
\overline{\V{\mathcal{W}}}_{j}^{n}(t)=\frac{1}{\D{x}\D{t}}\int_{n\D{t}}^{(n+1)\D{t}}\int_{(j-1)\D{x}}^{j\D{x}}\V{\mathcal{W}}(x,t)dxdt,\label{W_xt_average}\end{equation}
which is a normal random variable with zero mean and variance $\left(\D{x}\D{t}\right)^{-1}$,
independent between different cells and time steps. Note that for
certain types of equations the dynamic structure factor may be white
in frequency as well. In this case, a pointwise-in-time discretization
is not appropriate and one can instead use a spatio-temporal average
as done for white noise in (\ref{W_xt_average}).

We will study the accuracy of explicit linear finite-volume schemes
for solving the SPDE (\ref{U_t_linear_general}). Rather generally,
such methods are specified by a linear recursion of the form\begin{equation}
\V{U}^{n+1}=\left(\M{I}+\M{L}\D{t}\right)\V{U}^{n}+\sqrt{\frac{\D{t}}{\D{x}}}\M{K}\V{W}^{n},\label{U_np1_general}\end{equation}
where $\M{L}$ and $\M{K}$ are consistent stencil discretizations
of the continuum differential operators $\M{\mathcal{L}}$ and $\M{\mathcal{K}}$
(note that $\M{L}$ and $\M{K}$ may involve powers of $\D{t}$ in
general). Here \begin{equation}
\V{W}^{n}=\left(\D{x}\D{t}\right)^{\frac{1}{2}}\overline{\V{\mathcal{W}}}^{n}\label{dimensionless_W_j}\end{equation}
is a vector of standard normal variables with mean zero and variance
one.

Without the random forcing, the deterministic equation $\V{\mathcal{U}}_{t}=\M{\mathcal{L}}\V{\mathcal{U}}$
and the associated discretization can be studied using classical tools
and notions of stability, consistency, and convergence. Under the
assumption that the discrete generator $\M{L}$ is dissipative, the
initial condition $\V{U}^{0}$ will be damped and the equilibrium
solution will simply be a constant. The addition of the random forcing,
however, leads to a non-trivial invariant measure (equilibrium distribution)
of $\V{U}^{n}$ determined by an interplay between the (discretized)
fluctuations and dissipation. Because of the dissipative nature of
the generator, any memory of the initial condition will eventually
disappear and the long time dynamics is guaranteed to follow an ergodic
trajectory that samples the unique invariant measure. In order to
characterize the accuracy of the stochastic integrator, we will analyze
how well the discrete invariant measure (equilibrium distribution)
reproduces the invariant measure of the continuum SPDE (this is a
form of \emph{weak convergence}). Note that due to ergodicity, ensemble
averages can either be computed by averaging the power spectrum of
the fields over multiple samples or averaging over time (after sufficiently
many initial equilibration steps). In the theory we will consider
the limit $n\rightarrow\infty$ and then average over different realizations
of the noise $\V{W}$ to obtain the discrete structure factors. In
numerical calculations, we perform temporal averaging.

Regardless of the details of the iteration (\ref{U_np1_general}),
$\V{W}^{n}$ will always be a Gaussian random vector generated anew
at each step $n$ using a random number generator. The discretized
field $\V{U}^{n}$ is therefore a linear combination of Gaussian variates
and it is therefore a Gaussian vector-valued stochastic process. In
particular, the invariant measure (equilibrium distribution) of $\V{U}^{n}$
is fully characterized by the covariance \begin{equation}
\M{C}_{j,j^{\prime},n}^{(\V{U})}=\lim_{N_{s}\rightarrow\infty}\left\langle \V{U}_{j}^{N_{s}}\left(\V{U}_{j^{\prime}}^{N_{s}+n}\right)^{\star}\right\rangle ,\label{C_U_j_jp}\end{equation}
which we would like to compare to the covariance of the continuum
Gaussian field $\M{C}_{\V{\mathcal{U}}}\left(t=n\D{t}\right)$ given
by (\ref{C_U_general_continuum}). This comparison is best done in
the Fourier domain by using the spatial discrete Fourier transform,
defined for a spatially-discrete field $\V{U}$ {[}for example, $\V{U}\equiv\V{U}^{n}$
or $\V{U}\equiv\V{U}(t)$] via \begin{align}
\V{U}_{j}= & \sum_{\ki\in\widehat{\mathcal{V}}_{d}}\widehat{\V{U}}_{\ki}e^{ij\D{k}}\label{Discrete_FT_inverse}\\
\widehat{\V{U}}_{\ki}= & \frac{1}{V}\sum_{j=0}^{N_{c}-1}\V{U}_{j+1}e^{-ij\D{k}}\D{x},\label{Discrete_FT_forward}\end{align}
where we have denoted the discrete \emph{dimensionless} wavenumber
$\D{k}=k\D{x}=2\pi\kappa/N_{c}$, and the wave index is now limited
to the first $N_{c}$ values, \[
\widehat{\mathcal{V}}_{d}=\left\{ k=2\pi\kappa/H\quad\left|\quad\right.0\leq\kappa<N_{c}\right\} \subset\mathcal{K}.\]
Since the fields are real-valued, there is a redundancy in the Fourier
coefficients $\widehat{\V{U}}_{\ki}$ because of the Hermitian symmetry
between $\kappa$ and $N_{c}-\kappa$ (essentially, the second half
of the wave indices correspond to negative $k$), and thus we will
only consider $0\leq\kappa\leq\left\lfloor N_{c}/2\right\rfloor $,
giving a (Nyquist) cutoff wavenumber $k_{max}\approx\pi/\D{x}$.

What we would like to compare is the Fourier coefficients of the numerical
approximation, $\widehat{\V{U}}_{\ki}^{n}$, with the Fourier coefficients
of the continuum solution, $\widehat{\V{\mathcal{U}}}_{\ki}(t=n\D{t})$.
The invariant measure of $\widehat{\V{U}}_{\ki}^{n}$ has zero mean
and is characterized by the covariance obtained from the spatial Fourier
transform of (\ref{C_U_j_jp}),\begin{equation}
\M{S}_{\ki,n}=V\lim_{N_{s}\rightarrow\infty}\left\langle \widehat{\V{U}}_{\ki}^{N_{s}}\left(\widehat{\V{U}}_{\ki}^{N_{s}+n}\right)^{\star}\right\rangle .\label{S_k_t_discrete_def}\end{equation}
From the definition of the discrete Fourier transform it follows that
for small $\D{k}$, i.e., smooth Fourier basis functions on the scale
of the discrete grid, $\widehat{\V{U}}_{\ki}(t)$ converges to the
Fourier coefficient $\widehat{\V{\mathcal{U}}}(k,t=n\D{t})$ of the
continuum field. Therefore, $\M{S}_{\ki,n}$ is the discrete equivalent
(numerical approximation) to the continuum structure factor $\M{\mathcal{S}}(k,t=n\D{t})$.
We define a discrete approximation to be \emph{weakly consistent}
if \[
\lim_{\D{x},\D{t}\rightarrow0}\M{S}_{\ki,n=\left\lfloor t/\D{t}\right\rfloor }=\M{\mathcal{S}}\left(k,t\right),\]
for any chosen $\ki\in\widehat{\mathcal{V}}$ and $t$. This means
that, given a sufficiently fine discretization, the numerical scheme
can accurately reproduce the structure factor for a desired wave index
and time lag. An alternative view is that a convergent scheme reproduces
the slow (compared to $\D{t}$) and large-scale (compared to $\D{x}$)
fluctuations, that is, it accurately reproduces the dynamic structure
factor $\M{\mathcal{S}}(k,\omega)$ for small $\D{k}=k\D{x}$ and
$\D{\omega}=\omega\D{t}$. Our goal here is to quantify this for several
numerical methods for solving stochastic conservation laws and optimize
the numerical schemes by tuning parameters to obtain the best possible
approximation to $\M{\mathcal{S}}(k,\omega)$ for small $k$ and $\omega$.

Much of our analysis will be focused on the \emph{discrete static
structure factor}\[
\M{S}_{\ki}=\M{S}_{\ki,0}=V\lim_{N_{s}\rightarrow\infty}\left\langle \widehat{\V{U}}_{\ki}^{N_{s}}\left(\widehat{\V{U}}_{\ki}^{N_{s}}\right)^{\star}\right\rangle .\]
Note that for a spatially-white field $\V{\mathcal{U}}(x)$, the finite-volume
averages $\V{U}_{j}$ are independent Gaussian variates with mean
zero and variance $\D{x}^{-1}$, and the discrete Fourier coefficients
$\widehat{\V{U}}_{\ki}$ are independent Gaussian variates with mean
zero and variance $V^{-1}$. As a measure of the accuracy of numerical
schemes for solving Eq. (\ref{U_t_linear_general}), we will compare
the discrete static structure factors $\M{S}_{\ki}$ with the continuum
prediction $\M{\mathcal{S}}(k)$, for all of the discrete wavenumbers
(i.e., pointwise in Fourier space). It is expected that any numerical
scheme will produce some artifacts at the largest wavenumbers because
of the strong corrections due to the discretization; however, small
wavenumbers ought to have much smaller errors because they evolve
over time scales and length scales much larger than the discretization
step sizes. Specifically, we propose to look at the series expansions
\[
\M{S}_{\ki}-\M{\mathcal{S}}(k)=O\left(\D{t}^{p_{1}}k^{p_{2}}\right)\]
and optimize the numerical schemes by maximizing the powers $p_{1}$
and $p_{2}$. Next we describe the general formalism used to obtain
explicit expressions for the discrete structure factors $\M{S}_{\ki}$
for a general explicit method, and then illustrate the formalism on
some simple examples, before attacking the more complex equations
of fluctuating hydrodynamics.

\subsection{Analysis of Linear Explicit Methods\label{sub:Analysis-of-Linear-Explicit-Methods}}

Regardless of the details of a particular scheme and the particular
linear SPDE being solved, at the end of the timestep a typical explicit
scheme makes a linear combination of the values in the neighboring
cells and random variates to produce an updated value,\begin{equation}
\V{U}_{j}^{n+1}=\V{U}_{j}^{n}+\sum_{\D{j}=-w_{D}}^{\D{j}=w_{D}}\M{\Phi}_{\D{j}}\V{U}_{j+\D{j}}^{n}+\sum_{\D{j}=-w_{S}}^{\D{j}=w_{S}}\M{\Psi}_{\D{j}}\V{W}_{j+\D{j}}^{n},\label{dU_linear_generic}\end{equation}
where $w_{D}$ and $w_{S}$ are the deterministic and stochastic stencil
widths. The particular forms of the matrices of coefficients $\M{\Phi}$
and $\M{\Psi}$ depend on the scheme, and will involve powers of $\D{t}$
and $\D{x}$. Here we assume that for each $n$ the random increment
$\V{W}^{n}$ is an independent vector of $N_{s}$ normal variates
with covariance $\M{C}_{\V{W}}=\left\langle \V{W}_{j}^{n}\left(\V{W}_{j}^{n}\right)^{\star}\right\rangle $
constant for all of the cells $j$ and thus wavenumbers, where $N_{s}$
is the total number of random numbers utilized per cell per stage.
Computer algebra systems can be used to obtain explicit formulas for
the matrices in (\ref{dU_linear_generic}); we have made extensive
use of Maple for the calculations presented in this paper.

Assuming a translation invariant scheme, the iteration (\ref{dU_linear_generic})
can easily be converted from real space to an iteration in Fourier
space,\begin{equation}
\widehat{\V{U}}_{\ki}^{n+1}=\widehat{\V{U}}_{\ki}^{n}+\sum_{\D{j}=-w_{D}}^{\D{j}=w_{D}}\M{\Phi}_{\D{j}}\widehat{\V{U}}_{\ki}^{n}\exp\left(i\D{j}\D{k}\right)+\sum_{\D{j}=-w_{S}}^{\D{j}=w_{S}}\M{\Psi}_{\D{j}}\widehat{\V{W}}_{\ki}^{n}\exp\left(i\D{j}\D{k}\right),\label{generic_dU_hat_linear}\end{equation}
where different wavenumbers are not coupled to each other. In general,
any linear explicit method can be represented in Fourier space as
a recursion of the form\begin{equation}
\widehat{\V{U}}_{\ki}^{n+1}=\M{M}_{\ki}\widehat{\V{U}}_{\ki}^{n}+\M{N}_{\ki}\widehat{\V{W}}_{\ki}^{n},\label{U_k_np1}\end{equation}
where the explicit form of the matrices $\M{M}_{\ki}$ and $\M{N}_{\ki}$
depend on the particular scheme and typically contain various powers
of $\sin\D{k}$, $\cos\D{k}$, and $\D{t}$, and $\M{C}_{\widehat{\V{W}}}=\left\langle \widehat{\V{W}}_{\ki}^{n}\left(\widehat{\V{W}}_{\ki}^{n}\right)^{\star}\right\rangle =N_{c}^{-1}\M{C}_{\M{W}}$.
By iterating this recurrence relation, we can easily obtain (assuming
$\widehat{\V{U}}_{\ki}^{0}=0$)\[
\widehat{\V{U}}_{\ki}^{n+1}=\sum_{l=0}^{n}\left(\M{M}_{\ki}\right)^{l}\M{N}_{\ki}\widehat{\V{W}}_{\ki}^{n-l},\]
from which we can calculate\[
\M{S}_{\ki}^{n}=V\left\langle \left(\widehat{\V{U}}_{\ki}^{n}\right)\left(\widehat{\V{U}}_{\ki}^{n}\right)^{\star}\right\rangle =\sum_{l=0}^{n-1}\left(\M{M}_{\ki}\right)^{l}\left(\D{x}\M{N}_{\ki}\M{C}_{\V{W}}\M{N}_{\ki}^{\star}\right)\left(\M{M}_{\ki}^{\star}\right)^{l}=\sum_{l=0}^{n-1}\left(\M{M}_{\ki}\right)^{l}\widetilde{\M{C}}\left(\M{M}_{\ki}^{\star}\right)^{l}.\]
In order to calculate this sum explicitly, we will use the following
identity\[
\M{M}_{\ki}\M{S}_{\ki}^{n}\M{M}_{\ki}^{\star}-\M{S}_{\ki}^{n}=\left(\M{M}_{\ki}\right)^{n}\widetilde{\M{C}}\left(\M{M}_{\ki}^{\star}\right)^{n}-\widetilde{\M{C}}\]
to obtain a linear system for the entries of the matrix $\M{S}_{\ki}^{n}$.
If the deterministic method is stable, which means that all eigenvalues
of the matrix $\M{M}_{\ki}$ are below unity for all wavenumbers,
then in the limit $n\rightarrow\infty$ the first term on the right
hand side will vanish, to give \begin{equation}
\M{M}_{\ki}\M{S}_{\ki}\M{M}_{\ki}^{\star}-\M{S}_{\ki}=-\D{x}\M{N}_{\ki}\M{C}_{\V{W}}\M{N}_{\ki}^{\star}.\label{S_k_equation_general}\end{equation}
This is a linear system of equations for the equilibrium static structure
factor produced by a given scheme, where the number of unknowns is
equal to the square of the number of variables (field components).
By simply deleting the subscripts $k$ one obtains a more general
but much larger linear system \citet{MultigridMC_Goodman} for the
real space equilibrium covariance of a snapshot of the discrete field
$\M{C}_{j,j^{\prime}}^{(\V{U})}=\M{C}_{j,j^{\prime},n=0}^{(\V{U})}$,
\[
\M{M}\M{C}_{\V{U}}\M{M}^{\star}-\M{C}_{\V{U}}=-\D{x}\M{N}\M{C}_{\V{W}}^{(N_{c})}\M{N}^{\star},\]
where $\M{C}_{\V{W}}^{(N_{c})}=\left\langle \V{W}^{n}\left(\V{W}^{n}\right)^{\star}\right\rangle $
is the covariance matrix of the random increments. Note that this
relation continues to hold even for schemes that are not translation
invariant such as generalizations to non-periodic boundary conditions;
however, the number of unknowns is now the square of the total number
of degrees of freedom so that explicit solutions will in general not
be possible. Based on standard wisdom for deterministic schemes, it
is expected that schemes that perform well under periodic boundary
conditions will also perform well in the presence of boundaries when
the discretization is suitably modified only near the boundaries.

A similar approach to the one illustrated above for the static structure
factor can be used to evaluate the discrete \emph{dynamic} structure
factor\[
\M{S}_{\ki,\wi}=\lim_{N_{s}\rightarrow\infty}V\left(N_{s}\D{t}\right)\left\langle \widehat{\V{U}}_{\ki,\wi}^{N_{s}}\left(\widehat{\V{U}}_{\ki,\wi}^{N_{s}}\right)^{\star}\right\rangle \]
from the time-discrete Fourier transform\[
\widehat{\V{U}}_{\ki,\wi}^{N_{s}}=\frac{1}{N_{s}}\sum_{l=0}^{N_{s}}\exp\left(-il\D{\omega}\right)\widehat{\V{U}}_{\ki}^{l},\]
where $\D{\omega}=\omega\D{t}$, and the frequency is less than the
Nyquist cutoff, $\omega\leq\omega_{max}=\pi/\D{t}$. The calculation
yields\begin{equation}
\M{S}_{\ki,\wi}=\left[\M{I}-\exp\left(-i\D{\omega}\right)\M{M}_{\ki}\right]^{-1}\left(\D{x}\D{t}\M{N}_{\ki}\M{C}_{\V{W}}\M{N}_{\ki}^{\star}\right)\left[\M{I}-\exp\left(i\D{\omega}\right)\M{M}_{\ki}^{\star}\right]^{-1}.\label{S_kw_equation_general}\end{equation}
Equation (\ref{S_kw_equation_general}) can be seen as discretized
forms of the continuum version (\ref{S_U_dyn_continuum}) in the limits
$\D{k}\rightarrow0$, $\D{t}\rightarrow0$ (the corresponding correlations
in the time-domain are given in Ref. \citet{MultigridMC_Goodman}).

Equations (\ref{S_k_equation_general}) and (\ref{S_kw_equation_general})
are the main result of this section and we have used it to obtain
explicit expressions for $\M{S}_{\ki}$ and $\M{S}_{\ki,\wi}$ for
several equations and schemes. In the next sections we will illustrate
the above formalism for several simple examples of stochastic conservation
laws.

\subsubsection{\label{SectionDiscreteFD}Discrete Fluctuation-Dissipation Balance}

Let us first consider the static structure factors for very small
time steps. In the limit $\D{t}\rightarrow0$, temporal terms of order
two or more can be ignored so that all time-integration methods behave
like an explicit first-order Euler iteration as in (\ref{U_np1_general}),\begin{equation}
\widehat{\V{U}}_{\ki}^{n+1}=\left(\M{I}+\D{t}\widehat{\M{L}}_{\ki}^{(0)}\right)\widehat{\V{U}}_{\ki}^{n}+\sqrt{\frac{\D{t}}{\D{x}}}\widehat{\M{K}}_{\ki}^{(0)}\widehat{\V{W}}_{\ki},\label{U_np1_fourier_general}\end{equation}
where $\M{L}^{(0)}=\M{L}\left(\D{t}=0\right)$ can be thought of as
the spatial discretization of the generator $\M{\mathcal{L}}$, and
$\M{K}^{(0)}=\M{K}\left(\D{t}=0\right)$ is the spatial discretization
of the filtering operator $\M{\mathcal{K}}$. Comparing to (\ref{U_k_np1})
we can directly identify $\M{M}_{\ki}=\M{I}+\D{t}\widehat{\M{L}}_{\ki}^{(0)}$
and $\M{N}_{\ki}=\sqrt{\frac{\D{t}}{\D{x}}}\widehat{\M{K}}_{\ki}^{(0)}$
and substitute these into Eq. (\ref{S_k_equation_general}). Keeping
only terms of order $\D{t}$ on both sides we obtain the condition\begin{equation}
\widehat{\M{L}}_{\ki}^{(0)}\M{S}_{\ki}^{(0)}+\M{S}_{\ki}^{(0)}\left(\widehat{\M{L}}_{\ki}^{(0)}\right)^{\star}=-\widehat{\M{K}}_{\ki}^{(0)}\M{C}_{\V{W}}\left(\widehat{\M{K}}_{\ki}^{(0)}\right)^{\star},\label{S_k_general_SODE}\end{equation}
where $\M{S}_{\ki}^{(0)}=\lim_{\D{t}\rightarrow0}\M{S}_{\ki}$ (see
also a related real-space derivation using Ito's calculus in Ref.
\citet{AMR_ReactionDiffusion_Atzberger}, as well as Section VIII
in Ref. \citet{MultigridMC_Goodman}). It can be shown that if $\widehat{\M{L}}_{\ki}^{(0)}$
is definite, Eq. (\ref{S_k_general_SODE}) has a unique solution.
Assuming that $\M{W}$ is as given in Eq. (\ref{dimensionless_W_j}),
i.e., that $\M{C}_{\V{W}}=\M{I}$, and that the spatial discretizations
of the generator and filter operators satisfy a \emph{discrete fluctuation-dissipation
balance} \begin{equation}
\widehat{\M{L}}_{\ki}^{(0)}+\left(\widehat{\M{L}}_{\ki}^{(0)}\right)^{\star}=-\widehat{\M{K}}_{\ki}^{(0)}\left(\widehat{\M{K}}_{\ki}^{(0)}\right)^{\star},\label{discrete_FD_balance}\end{equation}
we see that $\M{S}_{\ki}^{(0)}=\M{I}$ is the solution to equation
(\ref{S_k_general_SODE}), that is, at equilibrium the discrete fields
are spatially-white. The discrete fluctuation-dissipation balance
condition can also be written in real space,\begin{equation}
\M{L}^{(0)}+\left(\M{L}^{(0)}\right)^{\star}=\M{K}^{(0)}\left(\M{K}^{(0)}\right)^{\star}.\label{discrete_FD_balance_real}\end{equation}
The condition (\ref{discrete_FD_balance_real}) is the discrete equivalent
of the continuum fluctuation-dissipation balance condition \citet{FluctuationDissipation_Kubo}
\begin{equation}
\M{\mathcal{L}}+\M{\mathcal{L}}^{\star}=-\M{\mathcal{K}}\M{\mathcal{K}}^{\star},\label{cont_FD_balance}\end{equation}
which ensures that $\M{\mathcal{S}}(k)=\M{I}$, i.e., that the invariant
measure of the SPDE is spatially-white. We observe that adding a skew
adjoint component to $\M{\mathcal{L}}$ does not alter the fluctuation-dissipation
balance above, as is the case with non-dissipative (advective) terms.
Numerous equations \citet{LebowitzHydroReview} modeling conservative
thermal systems satisfy condition (\ref{cont_FD_balance}), including
the linearized LLNS equations (with some additional prefactors). In
essence, the fluctuations injected at all scales by the spatially
white forcing $\M{\mathcal{W}}$ are filtered by $\M{\mathcal{K}}$
and then dissipated by $\M{\mathcal{L}}$ at just equal rates.

\section{\label{SectionLinearGeneral}Linear Stochastic Conservation Laws}

The remainder of this paper is devoted to the study of the accuracy
of finite-volume methods for solving linear stochastic PDEs in conservation
form, \begin{equation}
\partial_{t}\V{\mathcal{U}}=-\grad\cdot\left[\left(\M{A}\V{\mathcal{U}}-\M{C}\grad\V{\mathcal{U}}\right)-\M{E}\V{\mathcal{W}}\right],\label{U_t_linear_flux}\end{equation}
where $\M{A}$, $\M{C}$ and $\M{E}$ are constants, and $\V{\mathcal{W}}$
is Gaussian spatio-temporal white noise. The white noise forcing and
its divergence here need to be interpreted in the (weak) sense of
distributions since they lack the regularity required for the classical
definitions\textbf{.} The linearization of the LLNS equations (\ref{LLNS_general})
leads to a system of the form (\ref{U_t_linear_flux}), as do a number
of other classical PDEs \citet{LebowitzHydroReview}, such as the
\emph{stochastic advection-diffusion equation}\begin{equation}
\partial_{t}T=-\V{a}\cdot\grad T+\mu\grad^{2}T+\sqrt{2\mu}\grad\cdot\V{\mathcal{W}},\label{stoch_adv_diff_SPDE}\end{equation}
where $T(\V{r},t)\equiv\M{U}(\V{r},t)$ is a scalar stochastic field,
$\M{A\equiv a}$ is the advective velocity, $\M{C}\equiv\mu\M{I}$,
$\mu>0$ is the diffusion coefficient, and $\M{E}\equiv\sqrt{2\mu}\M{I}$.
The simplest case is the \emph{stochastic heat equation}, obtained
by taking $\V{a}=\V{0}$.

A key feature the type of system considered here is that the noise
is intrinsic to the system and appears in the flux as opposed to commonly
treated systems that include an external stochastic forcing term,
such as the form of a stochastic heat equation considered in Ref.
\citet{StochHeatEq_Weak}. Since white noise is more regular than
the spatial derivative of white noise, external noise leads to more
regular equilibrium fields (e.g., continuous functions in one dimension).
Intrinsic noise, on the other hand, leads to very irregular equilibrium
fields. Notationally, it is convenient to write (\ref{U_t_linear_flux})
as,\begin{equation}
\partial_{t}\V{\mathcal{U}}=-\M{\mathcal{D}}\left(\M{A}\V{\mathcal{U}}-\M{C}\M{\mathcal{G}}\V{\mathcal{U}}-\M{E}\V{\mathcal{W}}\right),\label{U_t_linear_flux_ACB}\end{equation}
defining the divergence $\M{\mathcal{D}}\equiv\grad\cdot$ and gradient
$\M{\mathcal{G}}\equiv\grad$ operators, $\M{\mathcal{D}}^{\star}=-\M{\mathcal{G}}$.
In the types of equations that appear in hydrodynamics, such as the
LLNS equations, the operator $\M{\mathcal{D}}\M{A}$ is skew-adjoint,
$\left(\M{\mathcal{D}}\M{A}\right)^{\star}=-\M{\mathcal{D}}\M{A}$
(hyperbolic or advective flux), $\M{C}\succeq\M{0}$ (dissipative
or diffusive flux), and $\M{E}\M{E}^{\star}=2\M{C}$, i.e., $\M{E}^{\star}=\left(2\M{C}\right)^{1/2}$.
Therefore, the generator $\M{\mathcal{L}}=-\M{\mathcal{D}}\M{A}+\M{\mathcal{D}}\M{C}\M{\mathcal{G}}=\left(\M{\mathcal{D}}\M{A}\right)^{\star}-\M{\mathcal{D}}\M{C}\M{\mathcal{D}}^{*}$
and filter $\M{\mathcal{K}}=\M{\mathcal{D}}\M{E}$ satisfy the fluctuation-dissipation
balance condition (\ref{cont_FD_balance}) and the equilibrium distribution
is spatially-white. Note that even though advection makes some of
the eigenvalues of $\M{\mathcal{L}}$ complex, the generator is dissipative
and (\ref{U_t_linear_flux}) has a unique invariant measure because
the real part of all of the eigenvalues of $\M{\mathcal{L}}$ is negative
except for the unique zero eigenvalue.

It is important to point out that discretizations of the continuum
operators do not necessarily satisfy the discrete fluctuation-dissipation
condition (\ref{discrete_FD_balance_real}). One way to ensure the
condition is satisfied is to discretize the diffusive components of
the generator $\M{L}_{D}=\M{D}\M{C}\M{G}$ and the filter $\M{K}=\M{D}\M{E}$
using a discrete divergence $\M{D}$ and discrete gradient $\M{G}$
so that the discrete fluctuation-dissipation balance condition $\M{L}_{D}+\M{L}_{D}^{\star}=-\M{K}\M{K}^{\star}$
holds. If, however, the discretization of the advective component
of the generator $\M{L}_{A}=-\M{D}\M{A}$ is not skew-adjoint, this
can perturb the balance (\ref{discrete_FD_balance}). Notably, various
upwinding methods lead to discretizations that are not skew-adjoint.
The correction to the structure factor $\M{S}_{\ki}^{(0)}=\M{I}+\D{\M{S}_{\ki}^{(0)}}$
due to a non-zero $\D{\M{L}_{A}}=\left(\M{L}_{A}+\M{L}_{A}^{\star}\right)/2$
can easily be obtained from Eq. (\ref{S_k_general_SODE}), and in
one dimension the result is simply \begin{equation}
\D{S_{\ki}^{(0)}}=-\frac{\D{L}_{k}^{(A)}}{L_{k}^{(D)}+\D{L}_{k}^{(A)}}.\label{upwinding_dS_k}\end{equation}
We will use centered differences for the advective generator in this
work, which ensures a skew-adjoint $\M{L}_{A}$, and our focus will
therefore be on satisfying the discrete fluctuation-dissipation balance
for the diffusive terms.

\subsection{Finite-Volume Numerical Schemes}

We consider here rather general finite-volume methods for solving
the linear SPDE (\ref{U_t_linear_flux}) in one dimension,\begin{equation}
\partial_{t}\V{\mathcal{U}}=-\frac{\partial}{\partial x}\left[\M{\mathcal{F}}(\V{\mathcal{U}})-\V{\mathcal{Z}}\right]=-\frac{\partial}{\partial x}\left[\left(\M{A}-\M{C}\frac{\partial}{\partial x}\right)\V{\mathcal{U}}-\M{E}\V{\mathcal{W}}\right]\label{U_t_linear_flux_1D}\end{equation}
with periodic boundaries, where we have denoted the modified (potentially
correlated) white noise flux with $\V{\mathcal{Z}}=\M{E}\V{\mathcal{W}}$.
As for classical finite-volume methods for the deterministic case,
we start from the PDE and integrate the left and right hand sides
over a given cell $j$ over a given time step $\D{t}$, and use integration
by parts to obtain the formally exact\begin{equation}
\V{U}_{j}^{n+1}=\V{U}_{j}^{n}-\frac{\D{t}}{\D{x}}\left(\V{F}_{j+\frac{1}{2}}-\V{F}_{j-\frac{1}{2}}\right)+\frac{\D{t}}{\D{x}}\left(\frac{1}{\sqrt{\Delta x\Delta t}}\right)\left(\V{Z}_{j+\frac{1}{2}}-\V{Z}_{j-\frac{1}{2}}\right),\label{generic_U_update_half}\end{equation}
where the \emph{deterministic discrete fluxes} $\V{F}$ and \emph{dimensionless
discrete stochastic fluxes} $\V{Z}$ are calculated on the boundaries
of the cells (points in one dimension, edges in two dimensions, and
faces in three dimensions), indexed here with half-integers. These
fluxes represent the total rate of transport through the interface
between two cells over a given finite time interval $\D{t}$, and
(\ref{generic_U_update_half}) is nothing more than a restatement
of conservation. The classical interpretation of pointwise evaluation
of the fluxes is not appropriate because white noise forcing lacks
the regularity of classical smooth forcing and cannot be represented
in a finite basis. Instead, just as we projected the fluctuating fields
using finite-volume averaging, we ought to project the fluxes to a
finite representation as well through spatio-temporal averaging, as
done in Eq. (\ref{W_xt_average}). For the purposes of our analysis,
one can simply think of the discrete fluxes as an approximation that
has the same spectral properties as the corresponding continuum Gaussian
fields over the wavevectors and frequencies represented by the finite
discretization.

The goal of numerical methods is to approximate the fluxes as best
as possible. In general, within each time step of a scheme there may
be $N_{st}$ stages or substeps; for example, in the classic MacCormack
method there is a predictor and a corrector stage ($N_{st}=2$), and
in the three-stage Runge-Kutta method of Williams \emph{et al.} \citet{Bell:07}
there are three stages ($N_{st}=3$). Each stage $0<s\leq N_{st}$
is of the conservative form (\ref{generic_U_update_half}),\begin{equation}
\V{U}_{j}^{n+\frac{s}{N_{st}}}=\sum_{s^{\prime}=0}^{s-1}\alpha_{s^{\prime}}^{(s)}\V{U}_{j}^{n+\frac{s^{\prime}}{N_{st}}}-\frac{\D{t}}{\D{x}}\left(\V{F}_{j+\frac{1}{2}}^{(s)}-\V{F}_{j-\frac{1}{2}}^{(s)}\right)+\frac{\D{t}^{1/2}}{\D{x}^{3/2}}\left(\V{Z}_{j+\frac{1}{2}}^{(s)}-\V{Z}_{j-\frac{1}{2}}^{(s)}\right),\label{generic_U_update_alpha}\end{equation}
where the $\alpha$'s are some coefficients, $\sum_{s^{\prime}=0}^{s-1}\alpha_{s^{\prime}}^{(s)}=1$,
and each of the stage fluxes are partial approximations of the continuum
flux. For the stochastic integrators we discuss here, the deterministic
fluxes are calculated the same way as they would be in the corresponding
deterministic scheme. In general, the stochastic fluxes $\V{Z}_{j+\frac{1}{2}}$
can be expressed in terms of independent unit normal variates $\V{W}_{j+\frac{1}{2}}$
that are sampled using a random number generator. The stochastic fluxes
in each stage may be the same, may be completely independent, or they
may have non-trivial correlations between stages.

Note that it is possible to avoid non-integer indices by re-indexing
the fluxes in Eq. (\ref{generic_U_update_half}) and writing it in
a form consistent with (\ref{dU_linear_generic}),\begin{equation}
\V{U}_{j}^{n+1}=\V{U}_{j}^{n}-\frac{\D{t}}{\D{x}}\left(\V{F}_{j}-\V{F}_{j-1}\right)+\frac{\D{t}^{1/2}}{\D{x}^{3/2}}\left(\V{Z}_{j}-\V{Z}_{j-1}\right).\label{generic_U_update}\end{equation}
However, when considering the order of accuracy of the stencils and
also fluctuation-dissipation balance in higher dimensions, it will
become important to keep in mind that the fluxes are evaluated on
the faces (edges or half-grid points) of the grid, and therefore we
will keep the half-integer indices. Note that for face-centered values,
such as fluxes, it is best to add a phase factor $\exp\left(i\D{k}/2\right)$
in the definition of the Fourier transform, even though such pure
phase shifts will not affect the correlation functions and structure
factors.

Before we analyze schemes for the complex LLNS equations, we present
an illustrative explicit calculation for the one-dimensional stochastic
heat equation.

\section{\label{sec:Section-Example:-Stochastic-Heat}Example: Stochastic
Heat Equation}

We now illustrate the general formalism presented in Section \ref{SectionLinearGeneral}
for the simple case of an Euler and predictor-corrector scheme for
solving the stochastic heat equation in one dimension,\begin{equation}
\upsilon_{t}=\mu\upsilon_{xx}+\sqrt{2\mu}\mathcal{W}_{x},\label{stochastic_diffusion_SPDE}\end{equation}
where $\upsilon\left(x,t\right)\equiv\mathcal{U}\left(x,t\right)$
is a scalar field and $\mu$ is the mass or heat diffusion coefficient.
The solution in the Fourier domain is trivial, giving\begin{equation}
S(k,\omega)=\frac{2\mu k^{2}}{\omega^{2}+\mu^{2}k^{4}},\mbox{ and }S(k)=1.\label{theoretical_S_kw_diffusion}\end{equation}

\subsection{\label{Section_S_k_diffusion}Static Structure Factor}

We first study a simple second-order spatial discretization of the
dissipative fluxes\[
F_{j+\frac{1}{2}}=\frac{\mu}{\D{x}}\left(u_{j+1}-u_{j}\right),\]
combined with an Euler integration in time, to give a simple numerical
method for solving the SPDE (\ref{stochastic_diffusion_SPDE}),\begin{equation}
u_{j}^{n+1}=u_{j}^{n}+\frac{\mu\D{t}}{\D{x}^{2}}\left(u_{j-1}^{n}-2u_{j}^{n}+u_{j+1}^{n}\right)+\sqrt{2\mu}\frac{\D{t}^{1/2}}{\D{x}^{3/2}}\left(W_{j+\frac{1}{2}}^{n}-W_{j-\frac{1}{2}}^{n}\right),\label{Euler_diffusion_iteration}\end{equation}
where $u\equiv U$ and the $W$'s are independent unit normal random
numbers with zero mean generated anew at every time step (here $N_{s}=N_{st}=1$).
From (\ref{Euler_diffusion_iteration}), we can extract the recursion
coefficients appearing in (\ref{U_k_np1}), \[
M_{\ki}=1+\beta(e^{-i\Delta k}-2+e^{i\Delta k})=1+2\beta\left(\cos\Delta k-1\right),\]
\[
N_{\ki}=\sqrt{2\mu}\frac{\D{t}^{1/2}}{\D{x}^{3/2}}\left(e^{i\Delta k/2}-e^{-i\Delta k/2}\right),\]
where \[
\beta=\frac{\mu\D{t}}{\D{x}^{2}}\]
denotes a dimensionless diffusive time step (ratio of the time step
to the diffusive CFL limit). Together with $C_{W}=1$, (\ref{S_k_equation_general})
becomes a scalar equation for the discrete structure factor,\[
(M_{k}M_{k}^{\star}-1)S_{k}=-\D{x}N_{k}N_{k}^{\star},\]
with dimensionless solution\begin{equation}
S_{\ki}=\frac{4\beta\left(1-\cos\D{k}\right)}{\left(1-M_{\ki}^{2}\right)}=\left[1+\beta\left(\cos\D{k}-1\right)\right]^{-1}.\label{S_k_diffusion}\end{equation}
The time-dependent result can also easily be derived from (\ref{S_kw_equation_general}),\[
S_{\ki}^{n}=\left(1-e^{-t/\tau}\right)S_{\ki},\mbox{ where }t=n\D{t}\]
where $\tau^{-1}=4\mu\left(\cos\D{k}-1\right)/\D{x}^{2}\approx2\mu k^{2}$
is the familiar relaxation time for wavenumber $k$, showing that
the smallest wavenumbers take a long time to reach the equilibrium
distribution.

\begin{figure}[tbph]
\begin{centering}
\includegraphics[width=0.49\textwidth]{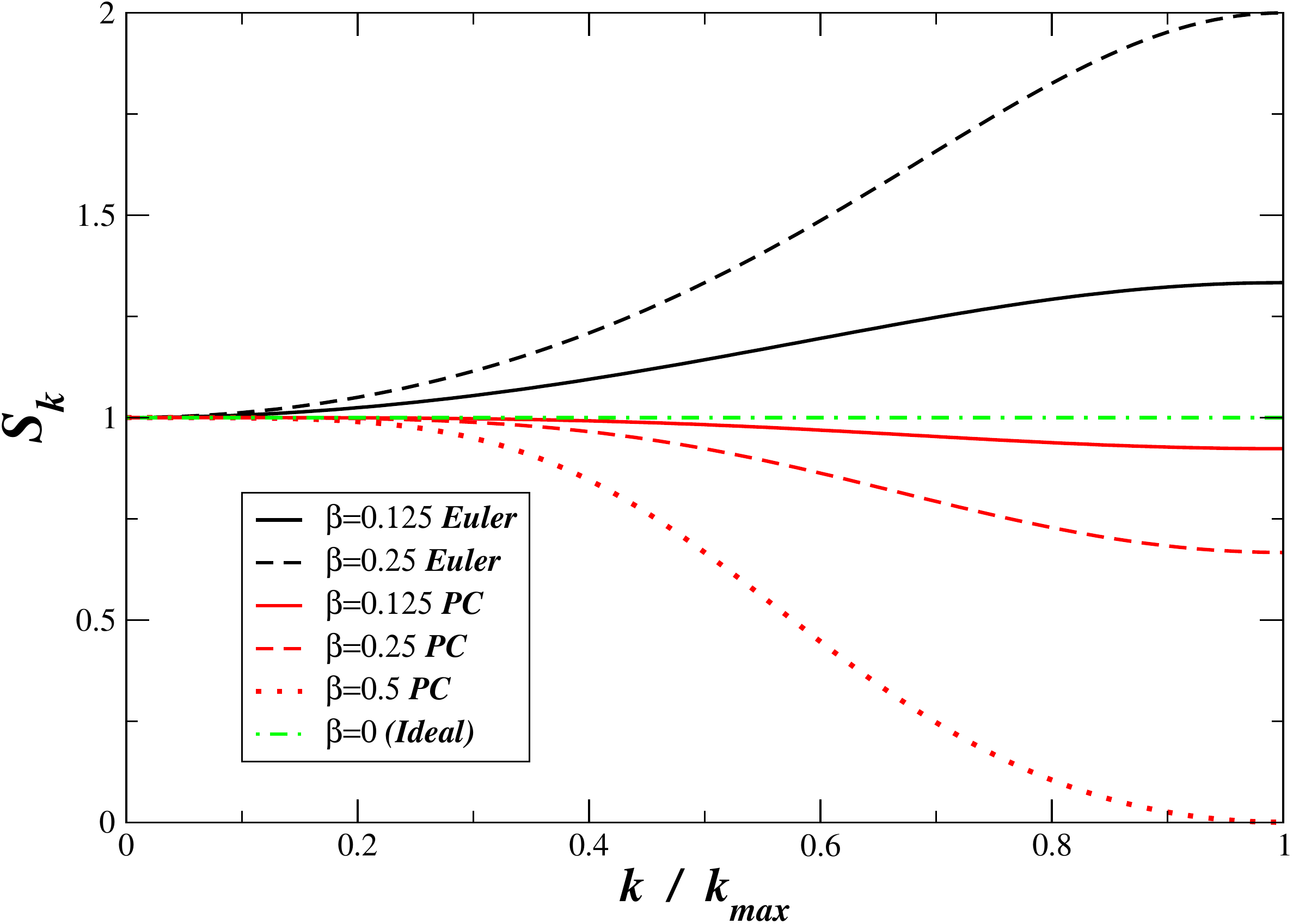}
\par\end{centering}

\caption{\label{fig:S_k_diffusion}An illustration of the discrete structure
factor $S_{k}$ for the Euler (\ref{Euler_diffusion_iteration}) {[}c.f.
Eq. (\ref{S_k_diffusion})] and predictor-corrector (\ref{PC_diffusion_iteration})
{[}c.f. Eq. (\ref{S_k_diffusion})] schemes for the stochastic heat
equation (\ref{stochastic_diffusion_SPDE}).}

\end{figure}

Equation (\ref{S_k_diffusion}) is a vivid illustration of the typical
result for schemes for stochastic transport equations based on finite
difference stencils, also shown in Fig. \ref{fig:S_k_diffusion}.
Firstly, we see that for small $k$ we have that $S_{\ki}\approx1+\beta\D{k}^{2}/2$,
showing that the smallest wavenumbers are correctly handled by the
discretization for any time step. Also, this shows that the error
in the structure factor is of order $\beta$, i.e., of order $\D{t}$,
as expected for the Euler scheme, whose weak order of convergence
is one for SODEs. Finally, it shows that the error grows quadratically
with $k$ (from symmetry arguments, only even powers will appear).
By looking at the largest wavenumber, $\D{k}_{max}=\pi$, we see that
$S_{\ki_{max}}=\left(1-2\beta\right)^{-1}$, from which we instantly
see the CFL stability condition $\beta<1/2$ , which guarantees that
the structure factor is finite and positive for all $0\leq k\leq\pi$.
Furthermore, we see that for $\beta\ll1$, the structure factor is
approximately unity for all wavenumbers. That is, a sufficiently small
step will indeed reproduce the proper equilibrium distribution.

By contrast, a two-stage predictor-corrector scheme for the diffusion
equation,\begin{align}
\tilde{u}_{j}^{n}= & u_{j}^{n}+\frac{\mu\D{t}}{\D{x}^{2}}\left(u_{j-1}^{n}-2u_{j}^{n}+u_{j+1}^{n}\right)+\sqrt{2\mu}\frac{\D{t}^{1/2}}{\D{x}^{3/2}}\left(W_{j+\frac{1}{2}}^{n}-W_{j-\frac{1}{2}}^{n}\right)\mbox{ (predictor)}\nonumber \\
u_{j}^{n+1}= & \frac{1}{2}\left[u_{j}^{n}+\tilde{u}_{j}^{n}+\frac{\mu\D{t}}{\D{x}^{2}}\left(\tilde{u}_{j-1}^{n}-2\tilde{u}_{j}^{n}+\tilde{u}_{j+1}^{n}\right)+\sqrt{2\mu}\frac{\D{t}^{1/2}}{\D{x}^{3/2}}\left(W_{j+\frac{1}{2}}^{n}-W_{j-\frac{1}{2}}^{n}\right)\right]\mbox{ (corrector),}\label{PC_diffusion_iteration}\end{align}
achieves much higher accuracy, namely, a structure factor that deviates
from unity by a higher order in both $\D{t}$ and $k$, \[
\mbox{PC-1RNG: }S_{\ki}\approx1-\beta^{2}\D{k}^{4}/4,\]
as illustrated in Fig. \ref{fig:S_k_diffusion}. We can also use different
stochastic fluxes in the predictor and the corrector schemes (i.e.,
use $N_{s}=2$ random numbers per cell per stage), with an added pre-factor
of $\sqrt{2}$ to compensate for the variance reduction of the averaging
between the two stages,\begin{align}
\tilde{u}_{j}^{n}= & u_{j}^{n}+\frac{\mu\D{t}}{\D{x}^{2}}\left(u_{j-1}^{n}-2u_{j}^{n}+u_{j+1}^{n}\right)+2\sqrt{\mu}\frac{\D{t}^{1/2}}{\D{x}^{3/2}}\left(W_{j+\frac{1}{2}}^{(n,P)}-W_{j-\frac{1}{2}}^{(n,P)}\right)\mbox{ (predictor)}\nonumber \\
u_{j}^{n+1}= & \frac{1}{2}\left[u_{j}^{n}+\tilde{u}_{j}^{n}+\frac{\mu\D{t}}{\D{x}^{2}}\left(\tilde{u}_{j-1}^{n}-2\tilde{u}_{j}^{n}+\tilde{u}_{j+1}^{n}\right)+2\sqrt{\mu}\frac{\D{t}^{1/2}}{\D{x}^{3/2}}\left(W_{j+\frac{1}{2}}^{(n,C)}-W_{j-\frac{1}{2}}^{(n,C)}\right)\right]\mbox{ (corrector).}\label{PC_diffusion_iteration2}\end{align}
For the scheme (\ref{PC_diffusion_iteration2}) the analysis reveals
an even greater spatio-temporal accuracy of the static structure factors,
namely, third order temporal accuracy\[
\mbox{PC-2RNG: }S_{\ki}\approx1+\beta^{3}\D{k}^{6}/8.\]
This illustrates the importance of the handling of the stochastic
fluxes in multi-stage algorithms, as we will come back to shortly.
Note that the analysis we presented here for explicit methods can
easily be extended to implicit and semi-implicit schemes as well,
as illustrated in Appendix \ref{SectionImplicit} for the Crank-Nicolson
method for the stochastic heat equation.

Previous studies \citet{Bell:07,FluctuatingHydro_Coveney} have measured
the accuracy of numerical schemes through the \emph{varianc}e of the
fields in real space, which, by Parseval's theorem, is related to
the integral of the structure factor over all wavenumbers. For the
Euler scheme (\ref{Euler_diffusion_iteration}) for the stochastic
heat equation this can be calculated analytically, \[
\sigma_{u}^{2}=\left\langle u_{j}^{2}\right\rangle -\left\langle u_{j}\right\rangle ^{2}=\D{x}^{-1}\left(1-2\beta\right)^{-1/2}\approx\D{x}^{-1}\left(1+\beta\right),\]
showing first-order temporal accuracy (in the weak sense). For the
predictor-corrector scheme (\ref{PC_diffusion_iteration}), on the
other hand, $\left(\sigma_{u}^{PC}\right)^{2}\approx\D{x}^{-1}\left(1-3\beta^{2}/2\right)$.
It is important to note, however, that using the variance as a measure
of accuracy of stochastic real-space integrators is both too rough
and also too stringent of a test. It does not give insights into how
well the equipartition is satisfied for the different modes, and,
at the same time, it requires that the structure factor be good even
for the highest wavenumbers, which is unreasonable to ask from a finite-stencil
scheme.

For pseudo-spectral methods, as studied for the incompressible fluctuating
Navier-Stokes equation in Ref. \citet{StochasticImmersedBoundary,StochasticImmersedBoundary_Theory},
one can modify the spectrum of the stochastic forcing so as to balance
the numerical stencil artifacts, and one can also use an (exact) exponential
temporal integrator in Fourier space to avoid the artifacts of time
stepping. However, for finite-volume schemes, a more reasonable approach
is to keep the stochastic fluxes uncorrelated between disjoint cells
(which is actually physical), and instead of looking at the variance,
focus on the accuracy of the static structure factor for small wavenumbers.
Specifically, basic schemes will typically have $\M{S}_{\ki}-1=O\left(\D{t}k^{2}\right)$,
while multi-step schemes will typically achieve $\M{S}_{\ki}-1=O\left(\D{t}^{2}k^{2}\right)$
or higher temporal order, or even $\M{S}_{\ki}-1=O\left(\D{t}^{2}k^{4}\right)$.

\subsection{Dynamic Structure Factor}

It is also constructive to study the full dynamic structure factor
for a given numerical scheme, especially for small wavenumbers and
low frequencies. This is significantly more involved in terms of analytical
calculations and the results are analytically more complicated, especially
for multi-stage methods and more complex equations. For the Euler
scheme (\ref{Euler_diffusion_iteration}) the solution to Eq. (\ref{S_kw_equation_general})
is\[
S_{\ki,\wi}=\frac{2\chi_{1}\chi_{2}^{-1}\mu k^{2}}{2\D{t}^{-2}\left(1-\cos\D{\omega}\right)+\chi_{1}^{2}\chi_{2}^{-1}\mu^{2}k^{4}},\]
where $\chi_{1}=2(1-\cos\D{k})/\D{k}^{2}$ and $\chi_{2}=1+2\beta\left(\cos\Delta k-1\right)$.
This shows that the dynamic structure factor does not converge to
the correct answer for all wavenumbers even in the limit $\D{t}\rightarrow0$,
namely \begin{equation}
\lim_{\beta\rightarrow0}S_{\ki,\wi}=\frac{2\chi_{1}\mu k^{2}}{\omega^{2}+\chi_{1}^{2}\mu^{2}k^{4}}.\label{S_kw_zero_dt}\end{equation}
For small $\D{k}$, $\chi_{1}\approx1-\D{k}^{2}/6$, and the numerical
result closely matches the theoretical result (\ref{theoretical_S_kw_diffusion}).
However, for finite wavenumbers the effective diffusion coefficient
is multiplied by a prefactor $\chi_{1}$, which represents the spatial
truncation error in the second-order approximation to the Laplacian.
For all of the time-integration schemes for the stochastic heat equation
discussed above, one can reduce the discrete dynamic structure factor
to a form\[
S_{\ki,\wi}=\frac{2\chi_{stoch}\mu k^{2}}{2\D{t}^{-2}\left(1-\cos\D{\omega}\right)+\chi_{det}^{2}\mu^{2}k^{4}},\]
where $\chi_{stoch}$ and $\chi_{det}$ depend on $\beta$ and $\D{k}$
and can be used to judge the accuracy of the scheme.

In this paper we focus on the static structure factors in order to
optimize the numerical schemes and then simply check numerically that
they also produce reasonably-accurate results for the dynamic structure
factors for small and intermediate wavenumbers and frequencies.

\subsection{\label{sub:Higher-Order-Differencing}Higher-Order Differencing}

Another interesting question is whether using a higher-order differencing
formula for the viscous fluxes improves upon the second-order formula
in the basic Euler scheme (\ref{Euler_diffusion_iteration}). For
example, a standard fourth order in space finite difference yields
the modified Euler scheme\begin{equation}
u_{j}^{n+1}=u_{j}^{n}+\frac{\mu\D{t}}{12\D{x}^{2}}\left(-u_{j-2}^{n}+16u_{j-1}^{n}-30u_{j}^{n}+16u_{j+1}^{n}-u_{j+2}^{n}\right)+\sqrt{2\mu}\frac{\D{t}^{1/2}}{\D{x}^{3/2}}\left(W_{j+\frac{1}{2}}-W_{j-\frac{1}{2}}\right).\label{Euler_diffusion_4th}\end{equation}
Repeating the previous calculation shows that\begin{equation}
\lim_{\beta\rightarrow0}S_{\ki}=6\left[7-\cos\D{k}\right]^{-1},\label{S_k_diffusion_4th}\end{equation}
demonstrating that the fluctuation-dissipation theorem is not satisfied
for this scheme at the discrete level even for infinitesimal time
steps. This is because the spatial discretization operators in (\ref{Euler_diffusion_4th})
do not satisfy the discrete fluctuation dissipation balance.

In order to obtain higher-order divergence and Laplacian stencils
that satisfy (\ref{discrete_FD_balance}) we can start from a higher
order divergence discretization $\M{D}$ and then simply calculate
the resulting discrete Laplacian $\M{L}=-\M{D}\M{D}^{\star}$. Here
$\M{D}$ should be a fourth-order (or higher) difference formula that
combines four face-centered values, two on each side of a given cell,
into an approximation to the derivative at the cell center. Conversely,
$\M{D}^{\star}$ combines the values from four cells, two on each
side of a given face, into an approximation to the derivative at the
face center. A standard fourth-order finite-difference stencil for
$\M{D}$ produces the \emph{higher-order Euler scheme},\begin{align}
u_{j}^{n+1} & =u_{j}^{n}+\frac{\mu\D{t}}{\D{x}^{2}}\left(\frac{1}{576}u_{j-3}^{n}-\frac{3}{32}u_{j-2}^{n}+\frac{87}{64}u_{j-1}^{n}-\frac{365}{144}u_{j}^{n}+\frac{87}{64}u_{j+1}^{n}-\frac{3}{32}u_{j+2}^{n}+\frac{1}{576}u_{j+3}^{n}\right)\nonumber \\
 & +\sqrt{2\mu}\frac{\D{t}^{1/2}}{\D{x}^{3/2}}\left(\frac{1}{24}W_{j-\frac{3}{2}}-\frac{9}{8}W_{j-\frac{1}{2}}+\frac{9}{8}W_{j+\frac{1}{2}}-\frac{1}{24}W_{j+\frac{3}{2}}\right),\label{Diffusion_higher_order}\end{align}
for which $S_{\ki}\approx1+\beta\D{k}^{2}/2$, which is the same leading-order
error as the basic Euler scheme (\ref{Euler_diffusion_iteration}).
On the other hand, the dynamic structure factor for small time steps
is as in Eq. (\ref{S_kw_zero_dt}) but now $\chi_{1}=(1-\cos\D{k})(13-\cos\D{k})/\left(72\D{k}^{2}\right)\approx1-3\D{k}^{4}/320$,
which shows the higher spatial order of the scheme.

Note that in (\ref{Diffusion_higher_order}) both the discretization
of the Laplacian and of the gradient are of higher spatial order than
in (\ref{Euler_diffusion_iteration}), however, the Laplacian operator
is not of the highest order possible for the given stencil width.
We will not use higher-order differencing for the diffusive fluxes
in this work in order to avoid large Laplacian stencils like the one
above. Rather, we will use the traditional second-order discretization
and focus on the time integration of the resulting system.

\subsection{Handling of Advection}

The analysis we illustrated here for the stochastic heat equation
can be directly applied to the scalar advection-diffusion equation
(\ref{stoch_adv_diff_SPDE}) in one dimension,\begin{equation}
\upsilon_{t}=-a\upsilon_{x}+\mu\upsilon_{xx}+\sqrt{2\mu}\mathcal{W}_{x}.\label{stoch_adv_diff_1D}\end{equation}
For example, a second-order centered difference discretization of
the advective term $-a\upsilon_{x}$ leads to the following explicit
Euler scheme\begin{equation}
u_{j}^{n+1}=u_{j}^{n}-\frac{\alpha}{2}\left(u_{j+1}^{n}-u_{j-1}^{n}\right)+\beta\left(u_{j-1}^{n}-2u_{j}^{n}+u_{j+1}^{n}\right)+\sqrt{2\mu}\frac{\D{t}^{1/2}}{\D{x}^{3/2}}\left(W_{j+\frac{1}{2}}^{n}-W_{j-\frac{1}{2}}^{n}\right),\label{Euler_diffadv_iteration}\end{equation}
where the dimensionless advective CFL number is\[
\alpha=\frac{a\D{t}}{\D{x}}=\beta r,\]
and $r=a\D{x}/\mu$ is the so-called cell Reynolds number and measures
the relative importance of advective and diffusive terms at the grid
scale. Note that this scheme is unconditionally unstable when $\mu=0$,
specifically, the stability condition is $\alpha^{2}/2\leq\beta\leq1/2$.

For the Euler method (\ref{Euler_diffadv_iteration}) the analysis
yields a structure factor\[
S_{\ki}\approx\frac{1}{1-\alpha r/2}+\frac{\left(1-r^{2}/4\right)}{2\left(1-\alpha r/2\right)^{2}}\beta\D{k}^{2},\]
showing that even the smallest wavenumbers have the wrong spectrum
for a finite time step when $\left|r\right|>0$, which is unacceptable
in practice since it means that even the slowly-evolving large-scale
fluctuations are not handled correctly. Adding an artificial diffusion
$\D{\mu}=\mu\left|r\right|/2$ to $\mu$ leads to an improved leading
order error,\[
S_{\ki}\approx1+\frac{\left(1-r^{2}/4\right)}{2}\beta\D{k}^{2}+O(\D{t}^{2}\D{k}^{2}).\]
It is well-known that adding such an artificial diffusion is equivalent
to upwinding the advective term and leads to much improved stability
for large $r$ as well%
\footnote{Note that for this particular type of upwinding the denominator in
Eq. (\ref{upwinding_dS_k}) vanishes identically and it can be shown
that the correct solution is $\D{S_{\ki}^{(0)}}=0$, however, this
is not necessarily true for other, higher order, upwind discretizations
of advection.%
}.

The second-order predictor-corrector time stepping scheme can be applied
when advection is included as well. If $\left|r\right|>0$ the leading
order errors are\begin{align}
\mbox{PC-1RNG: } & S_{\ki}\approx1-\frac{\alpha^{2}}{4}\left(1-\frac{r\alpha}{2}\right)\D{k}^{2}\nonumber \\
\mbox{PC-2RNG: } & S_{\ki}\approx1-\frac{r\alpha^{3}}{8}\D{k}^{2},\label{PC_advection_results}\end{align}
showing that PC-2RNG gives a more accurate discrete structure factor
than PC-1RNG even if advection is included as well. Note that the
predictor-corrector method is unconditionally unstable when $\mu=0$.
In Section \ref{Section_RK3} we analyze a three-stage Runge-Kutta
scheme that has a small leading order error in $S_{k}$ but is also
stable when $\alpha<1$ even if $\mu=0$.

\section{\label{sec:Section-LLNS-Equations-1D}LLNS Equations in One Dimension}

In this section, we will consider the linearized LLNS system (\ref{LLNS_linear_ideal})
for a mono-atomic ideal gas in one spatial dimension, that is, where
symmetry dictates variability along only the $x$ axis. As explained
in the Introduction, focusing on an ideal gas simply fixes the values
of certain coefficients and thus simplifies the algebra, without limiting
the generality of our analysis. We will arbitrarily choose the number
of degrees of freedom per particle to be $d_{f}=1$, even though in
most cases of physical interest $d_{f}=3$ is appropriate; this merely
changes some of the constant coefficients and does not affect our
discussion. Explicitly, the one-dimensional linearized LLNS equations
are\begin{equation}
\left[\begin{array}{c}
\partial_{t}\rho\\
\partial_{t}v\\
\partial_{t}T\end{array}\right]=-\frac{\partial}{\partial x}\left[\begin{array}{c}
\rho_{0}v+\rho v_{0}\\
c_{0}^{2}\rho_{0}^{-1}\rho+c_{0}^{2}T_{0}^{-1}T+v_{0}v\\
c_{0}^{2}c_{v}^{-1}v+Tv_{0}\end{array}\right]+\frac{\partial}{\partial x}\left[\begin{array}{c}
0\\
\rho_{0}^{-1}\eta_{0}v_{x}\\
\rho_{0}^{-1}c_{v}^{-1}\mu_{0}T_{x}\end{array}\right]+\frac{\partial}{\partial x}\left[\begin{array}{c}
0\\
\rho_{0}^{-1}\Sigma\\
\rho_{0}^{-1}c_{v}^{-1}\Xi\end{array}\right],\label{LLNS_1D_ideal}\end{equation}
where the covariance matrices of the stochastic fluxes are $C_{\Sigma}=2\eta_{0}k_{B}T_{0}$
and $C_{\Xi}=2\mu_{0}k_{B}T_{0}^{2}$. In Fourier space the flux becomes\[
\widehat{\M{F}}=\left[\begin{array}{ccc}
v_{0} & \rho_{0} & 0\\
\rho_{0}^{-1}c_{0}^{2} & \left(v_{0}-ik\rho_{0}^{-1}\eta_{0}\right) & T_{0}^{-1}c_{0}^{2}\\
0 & c_{0}^{2}c_{v}^{-1} & \left(v_{0}-ik\rho_{0}^{-1}c_{v}^{-1}\mu_{0}\right)\end{array}\right],\]
which through Eqs. (\ref{S_U_dyn_continuum}) and (\ref{S_U_static_continuum})
(or, equivalently, Eq. (\ref{S_k_general_SODE})) gives static structure
factors that are independent of $k$,\begin{equation}
\M{S}(k)=\left[\begin{array}{ccc}
\rho_{0}c_{0}^{-2}k_{B}T_{0} & 0 & 0\\
0 & \rho_{0}^{-1}k_{B}T_{0} & 0\\
0 & 0 & \rho_{0}^{-1}c_{v}^{-1}k_{B}T_{0}^{2}\end{array}\right].\label{S_U_continuum}\end{equation}
Therefore, the invariant distribution for the spatial fluctuating
fields is white noise, uncorrelated among the different primitive
variables, and with variances given in Eq. (\ref{S_U_continuum}).
This is in agreement with predictions of statistical mechanics, and
how Landau and Lifshitz obtained the form of the stochastic fluxes.
Note that in the incompressible limit, $c_{0}\rightarrow\infty$,
the density fluctuations diminish, but the velocity and temperature
fluctuations are independent of $c_{0}$.

In this section we will calculate the discrete structure factor for
several finite-volume approximations to (\ref{LLNS_1D_ideal}). From
the diagonal elements of $\M{S}_{\ki}$ we can directly obtain the
non-dimensionalized static structure factors for the three primitive
variables, for example,\[
S_{\ki}^{(\rho)}=\frac{V}{\rho_{0}c_{0}^{-2}k_{B}T_{0}}\left\langle \hat{\rho}_{\ki}\hat{\rho}_{\ki}^{\star}\right\rangle ,\]
which for a perfect scheme would be unity for all wavevectors. Similarly,
the off-diagonal or cross elements, such as for example\[
S_{\ki}^{(\rho,v)}=\frac{V}{\sqrt{\left(\rho_{0}c_{0}^{-2}k_{B}T_{0}\right)\left(\rho_{0}^{-1}k_{B}T_{0}\right)}}\left\langle \hat{\rho}_{k}\hat{v}_{\ki}^{\star}\right\rangle ,\]
would all vanish for all wavevectors for a perfect scheme. Our goal
will be to quantify the deviations from {}``perfect'' for several
methods, as a function of the discretization parameters $\D{x}$ and
$\D{t}$.

\subsection{\label{Section_RK3}Third-order Runge-Kutta (RK3) Scheme}

When designing numerical schemes to integrate the full LLNS system,
it seems most appropriate to base the scheme on well-known robust
deterministic methods, and modify the deterministic methods by simply
adding a stochastic component to the fluxes, in addition to the usual
deterministic component. With such an approach, at least we can be
confident that in the case of weak noise the solver will be robust
and thus we will not compromise the fluid solver just to accommodate
the fluctuations.

A well-known approach to solving PDEs in conservation form\[
\partial_{t}\V{\mathcal{U}}=-\grad\cdot\left[\M{\mathcal{F}}(\V{\mathcal{U}})\right]=-\grad\cdot\left[\M{\mathcal{F}}_{H}(\V{\mathcal{U}})+\M{\mathcal{F}}_{D}(\grad\V{\mathcal{U}})\right]\]
is to use the \emph{method of lines} to decouple the spatial and temporal
discretizations. We will focus on one dimension first for notational
simplicity. In the method of lines, a finite-volume spatial discretization
is applied to the obtain a system of stochastic differential equations
for the discretized fields:\begin{align}
\frac{d\V{U}_{j}}{dt} & =-\D{x}^{-1}\left[\V{F}_{j+\frac{1}{2}}(\V{U})-\V{F}_{j-\frac{1}{2}}(\V{U})\right]=\nonumber \\
 & =-\D{x}^{-1}\left[\V{F}_{H}(\V{U}_{j+\frac{1}{2}})-\V{F}_{H}(\V{U}_{j-\frac{1}{2}})\right]-\D{x}^{-1}\left[\V{F}_{D}(\grad_{j+\frac{1}{2}}\V{U})-\V{F}_{D}(\grad_{j-\frac{1}{2}}\V{U})\right],\label{method_of_lines_1D}\end{align}
where $\V{U}_{j+\frac{1}{2}}$ are face-centered values of the fields
that are calculated from the cell-centered values $\V{U}_{j}$, and
$\grad_{j+\frac{1}{2}}$ is a cell-to-face discretization of the gradient
operator. Any classical temporal integrator can be applied to the
resulting system of SODEs. It is well known that the Euler and Heun
(two-step second-order Runge-Kutta) methods are unconditionally unstable
for hyperbolic equations. In Ref. \citet{Bell:07}, an algorithm for
the solution of the LLNS system of equations (\ref{LLNS_general})
was proposed, which is based on the three-stage, low-storage TVD Runge-Kutta
(RK3) scheme of Gottlieb and Shu \citet{Gottlieb:98}. The RK3 scheme
is the simplest TVD RK discretization for the deterministic compressible
Navier-Stokes equations that is stable even in the inviscid limit,
with the omission of slope-limiting. Here we adopt the same basic
scheme and investigate optimal ways of evaluating the stochastic flux.

In the RK3 scheme, the hyperbolic component of the face flux $\V{F}_{H}$
is calculated by a cubic interpolation of $\V{U}$ from the cell centers
to the faces using an interpolation formula borrowed from the PPM
method \citet{PPM_Collela},\begin{equation}
\V{U}_{j+\frac{1}{2}}=\frac{7}{12}\left(\V{U}_{j}+\V{U}_{j+1}\right)-\frac{1}{12}\left(\V{U}_{j-1}+\V{U}_{j+2}\right),\label{PPM_interpolation}\end{equation}
and then directly evaluating the hyperbolic flux from the interpolated
values. In Refs. \citet{Bell:07,Bell:09} a modified interpolation
is proposed that preserves variances; however, our analytical calculations
indicate that this type of interpolation artificially increases the
structure factor for intermediate wavenumbers in order to compensate
for the errors at larger wavenumbers. Note that for the full non-linear
equations, the conserved quantities are interpolated and then primitive
face variables are calculated from those. For the linearized equations
it does not matter and it is simpler to work exclusively with primitive
variables. In the RK3 method, the diffusive components of the fluxes
$\V{F}_{D}$ are calculated using classical face-centered second-order
centered stencils to evaluate the gradients of the fields at the cell
faces. Stochastic fluxes $\V{Z}_{j+\frac{1}{2}}$ are also generated
at the faces of the grid using a standard random number generator
(RNG). These stochastic fluxes are generated independently for velocity
and temperature, and are zero for density,\[
\V{Z}_{j+\frac{1}{2}}^{(RNG)}=\left[\begin{array}{c}
0\\
\rho_{0}^{-1}\left(2\eta_{0}k_{B}T_{0}\right)^{\frac{1}{2}}W_{j+\frac{1}{2}}^{(1)}\\
\rho_{0}^{-1}c_{v}^{-1}\left(2\mu_{0}k_{B}T_{0}^{2}\right)^{\frac{1}{2}}W_{j+\frac{1}{2}}^{(2)}\end{array}\right],\]
where $W_{j+\frac{1}{2}}^{(1/2)}$ denotes a normal variate with zero
mean and unit variance.

For each stage of the RK3 scheme, a total cell increment is calculated
as

\[
\D{\V{U}}_{j}(\V{U},\V{W})=-\frac{\D{t}}{\D{x}}\left[\V{F}_{j+\frac{1}{2}}(\V{U})-\V{F}_{j-\frac{1}{2}}(\V{U})\right]+\frac{\D{t}^{1/2}}{\D{x}^{3/2}}\left(\V{Z}_{j+\frac{1}{2}}-\V{Z}_{j-\frac{1}{2}}\right).\]
Each time step of the RK3 algorithm is composed of three stages\begin{align}
\V{U}_{j}^{n+\frac{1}{3}}= & \V{U}_{j}^{n}+\D{\V{U}}_{j}(\V{U}^{n},\V{W}_{1})\mbox{ (estimate at }t=(n+1)\D{t}\mbox{ )}\nonumber \\
\V{U}_{j}^{n+\frac{2}{3}}= & \frac{3}{4}\V{U}_{j}^{n}+\frac{1}{4}\left[\V{U}_{j}^{n+\frac{1}{3}}+\D{\V{U}}_{j}(\V{U}_{j}^{n+\frac{1}{3}},\V{W}_{2})\right]\mbox{ (estimate at }t=(n+\frac{1}{2})\D{t}\mbox{ )}\nonumber \\
\V{U}_{j}^{n+1}= & \frac{1}{3}\V{U}_{j}^{n}+\frac{2}{3}\left[\V{U}_{j}^{n+\frac{2}{3}}+\D{\V{U}}_{j}(\V{U}^{n+\frac{2}{3}},\V{W}_{3})\right],\label{RK3_explicit_stages}\end{align}
where for now we have not assumed anything about how the stochastic
fluxes between different stages, $\V{W}_{1}$, $\V{W}_{2}$ and $\V{W}_{3}$,
are related to each other. The relevant dimensionless parameters that
measure the ratio of the time step to the CFL stability limits are\begin{align*}
\alpha= & \frac{c_{0}\D{t}}{\D{x}}\\
\beta= & \frac{\eta_{0}\D{t}}{\rho_{0}\D{x}^{2}}=\frac{\alpha}{r}\\
\beta_{T}= & \frac{\mu_{0}\D{t}}{\rho_{0}c_{v}\D{x}^{2}}=\frac{1}{\mbox{Pr}}\frac{\alpha}{r}=\frac{\alpha}{p},\end{align*}
where $r=c_{0}\rho_{0}\D{x}/\eta_{0}$ is the cell Reynolds number
and measures the relative importance of acoustic and viscous terms
at the grid scale (we have assumed a low Mach number flow, i.e., $\left|v_{0}\right|\ll c_{0}$),
and $\mbox{Pr}=\eta_{0}c_{v}/\mu_{0}$ is the Prandtl number of the
fluid. For low-density gases, $r$ and $p=r\mbox{Pr}$ can be close
to or smaller than one, however, for dense fluids sound dominates
and $r>1$ and $p>1$ for all reasonable $\D{x}$ (essentially, $\D{x}>\lambda$,
where $\lambda$ is the mean free path). In practice, in order to
fully resolve viscous scales, one should keep both $r$ and $p$ reasonably
small.

\subsection{Evaluation of the Stochastic Fluxes}

In the original RK3 algorithm \citet{Bell:07}, a different stochastic
flux is generated in each stage, that is, $\V{W}_{s}=\sqrt{2}\V{W}_{RNG}^{(s)}$,
$s=1\ldots3$. The additional prefactor $\sqrt{2}$ is added because
the averaging between the three stages reduces the variance of the
overall stochastic flux. One can also use different weights for each
of the three stochastic fluxes, i.e., $\V{W}_{s}=w_{s}\V{W}_{RNG}^{(s)}$.
Another option is to simply use the same stochastic flux $\V{W}_{RNG}^{(0)}$
in all three stages, that is, $\V{W}_{s}=\V{W}_{RNG}^{(0)}$. A further
option is to use the same random flux $\V{W}_{RNG}^{(0)}$ in all
three stages, but put in different weights in each stage, i.e., $\V{W}_{s}=w_{s}\V{W}_{RNG}^{(0)}.$
Our goal is to find out which approach is optimal. For this purpose,
we can generally assume that the three random fluxes are different,
to obtain a total of six random numbers per cell per step, and use
the formalism developed in Section \ref{sec:Explicit-Methods} with
$N_{s}=6$ to express the structure factor in terms of the $6\times6$
covariance matrix of the random variates. This calculation is too
tedious even for a computer algebra system, and we therefore first
study the simple advection-diffusion equation (\ref{stoch_adv_diff_SPDE})
in order to gain some insight.

\subsubsection{\label{SectionAdvectionDiffusion}Advection-Diffusion Equation}

The RK3 method can be directly applied to the scalar advection-diffusion
equation in one dimension (\ref{stoch_adv_diff_1D}). Experience with
deterministic solvers suggests that a numerical scheme that performs
well on this type of model equation is likely to perform well on the
full system (\ref{LLNS_general}) when viscous effects are fully resolved.
Here we use the PPM-interpolation based discretization of the hyperbolic
flux given in Eq. (\ref{PPM_interpolation}), which leads to a standard
fourth-order centered difference approximation to the first derivative
$\upsilon_{x}$ \citet{bao2003hos} (in Fourier space the relative
error in the hyperbolic flux is of order $O(\D{k}^{4})$), and thus
justifies our choice for the interpolation. We discretize the gradient
used in calculating the diffusive fluxes using the second-order centered
difference\[
\grad_{j+\frac{1}{2}}u=\frac{u_{j+1}-u_{j}}{\Delta x},\]
which leads to the standard second-order centered difference approximation
to the second derivative $\upsilon_{xx}$ (the challenges with using
the standard fourth-order centered difference approximation to $\upsilon_{xx}$
\citet{bao2003hos} are discussed in Section \ref{sub:Higher-Order-Differencing}).
The stencil widths in Eq. (\ref{dU_linear_generic}) are $w_{D}=6$
(three stages with stencil width two each) and $w_{S}=4$, and there
are $N_{s}=3$ random numbers per cell per step (one per stage), with
a general $3\times3$ covariance matrix $\M{C}_{\V{W}}$. Equation
(\ref{S_k_equation_general}) can then be solved to obtain the static
structure factor for any wavenumber, however, these expressions are
too complex to be useful for analysis. Instead, we perform an expansion
of both sides of (\ref{S_k_equation_general}) for small $k$ and
thus focus on the behavior of the static structure factors for small
wavenumbers and small time steps.

As a first condition on $\M{C}_{\V{W}}$, we have the weak consistency
requirement $S_{\ki=0}=1$. With this condition satisfied, the method
satisfies the discrete fluctuation-dissipation balance in the limit
$\D{t}\rightarrow0$ since the discretization of the divergence is
the negative adjoint of the discretization of the gradient. A second
condition is obtained by equating the coefficient in front of the
leading-order error term in $S_{\ki}$, of order $\alpha\D{k}^{2}$,
to zero; where the advective dimensionless CFL number is $\alpha=a\Delta t/\Delta x$.
It turns out that this also makes the term of order $\alpha\D{k}^{4}$
vanish. A third condition is obtained by equating the coefficient
in front of the next-order error term of order $\alpha^{2}\D{k}^{2}$
to zero. Finally, a fourth condition equates the coefficient in front
of $\alpha^{2}\D{k}^{4}$ to zero. For this three-stage method, it
is not possible to make the terms with higher powers of $\alpha$
vanish identically for any choice of $\M{C}_{\V{W}}$. No additional
conditions are obtained by looking at terms with powers of the diffusive
CFL number $\beta=\mu\D{t}/\D{x}^{2}$ since, as it turns out, the
accuracy is always limited by the hyperbolic fluxes.

The various ways of generating the stochastic fluxes can now be compared
by investigating how many of these conditions are satisfied. It turns
out that only the first condition is satisfied if we use a different
independently-generated stochastic flux in each stage (one can satisfy
one more condition by using different weights for the three independent
stochastic fluxes). The second condition is satisfied if we use the
same stochastic flux in all stages with a unit weight, i.e., $\V{W}_{s}=w_{s}\V{W}_{RNG}^{(0)}$
with $w_{1}=w_{2}=w_{3}=1$. Armed with the freedom to put a different
weight for this flux in each of the stages, we can satisfy the third
condition as well if we use\begin{equation}
w_{1}=\frac{3}{4},\mbox{ }w_{2}=\frac{3}{2},\mbox{ }w_{3}=\frac{15}{16},\label{RK3_magic}\end{equation}
which gives a structure factor\[
S_{\ki}=1-\frac{r}{24}\alpha^{3}\D{k}^{2}-\frac{1}{6r^{2}}\alpha^{2}\D{k}^{4}+\mbox{h.o.t.}\]

If we are willing to increase the cost of each step and generate two
random numbers per cell per step, we can satisfy the fourth condition
as well. For this purpose, we look for a covariance matrix $\M{C}_{\M{W}}$
that satisfies the four conditions and is also positive semi-definite
and has a rank of two, i.e., has a smallest eigenvalue of zero. A
solution to these equations gives the following method for evaluating
the stochastic fluxes in the three stages\begin{align}
\V{W}_{1}= & \V{W}_{RNG}^{(A)}-\sqrt{3}\V{W}_{RNG}^{(B)}\nonumber \\
\V{W}_{2}= & \V{W}_{RNG}^{(A)}+\sqrt{3}\V{W}_{RNG}^{(B)}\nonumber \\
\V{W}_{3}= & \V{W}_{RNG}^{(A)},\label{RK3_optimal}\end{align}
where $\V{W}_{RNG}^{(A)}$ and $\V{W}_{RNG}^{(B)}$ are two independent
random vectors that need to be generated and stored during each RK3
step. This approach produces a structure factor\[
S_{\ki}=1-\frac{r}{24}\alpha^{3}\D{k}^{2}-\frac{24+r^{2}}{288r}\alpha^{3}\D{k}^{4}+\mbox{h.o.t.}\]
We will refer to the RK3 scheme that uses one random flux per step
and the weights in (\ref{RK3_magic}) as the \emph{RK3-1RNG scheme},
and to the RK3 scheme with two random fluxes per step as given in
(\ref{RK3_optimal}) as the \emph{RK3-2RNG scheme.}

It is important to point out that for the MacCormack method, which
is equivalent to the Lax-Wendroff method for the advection-diffusion
equation, the leading-order errors are of order $\alpha\D{k}^{2}$.
This is much worse than for the stochastic heat equation (see Section
\ref{Section_S_k_diffusion}) even though the MacCormack scheme is
a predictor-corrector method. This is because of the low-order handling
of advective fluxes used in the MacCormack method to stabilize the
two-stage Runge-Kutta time integrator.

\subsection{Results for LLNS equations in One Dimension}

\begin{figure}[tbph]
\begin{centering}
\includegraphics[width=0.45\textwidth]{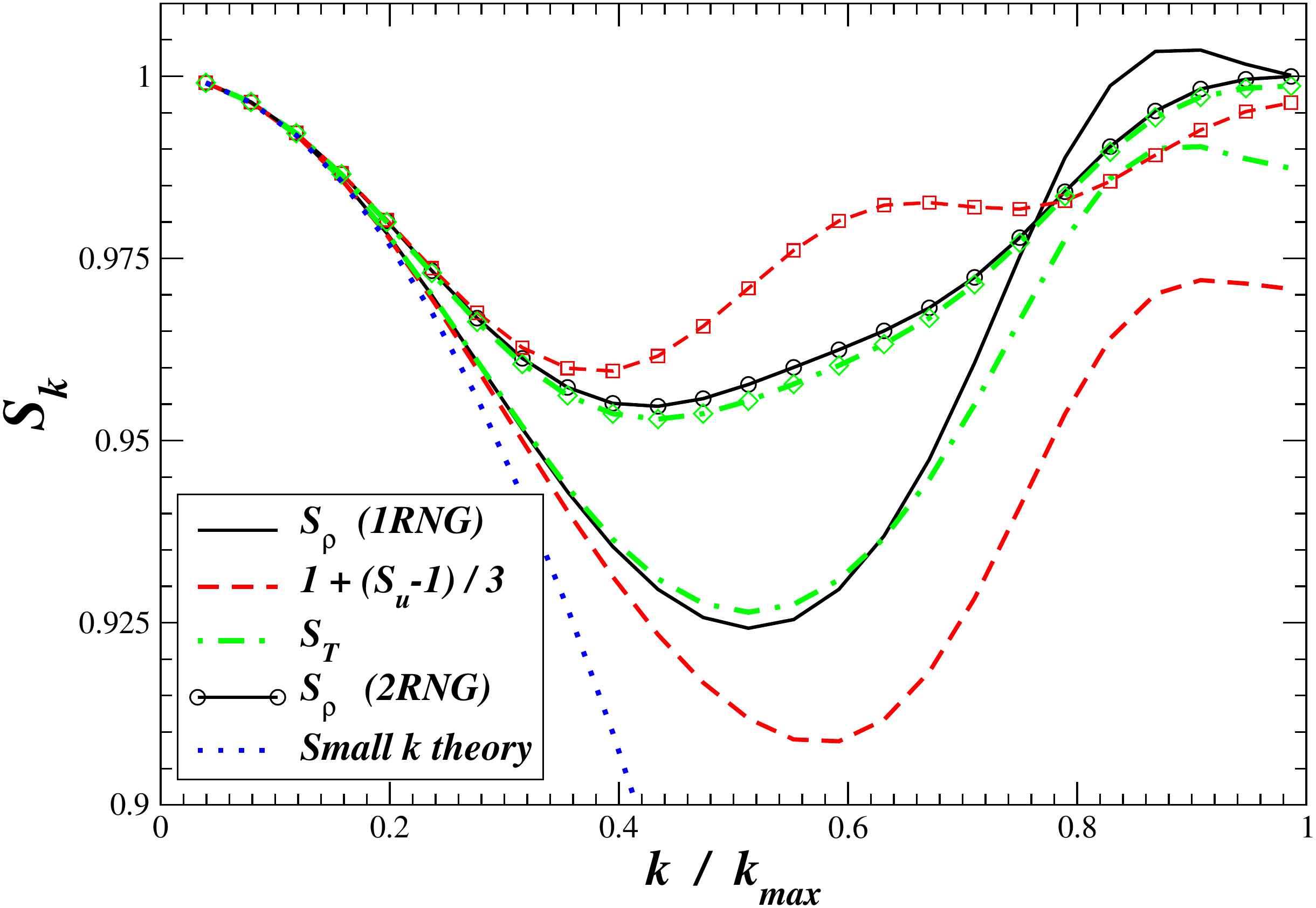}\includegraphics[width=0.45\textwidth]{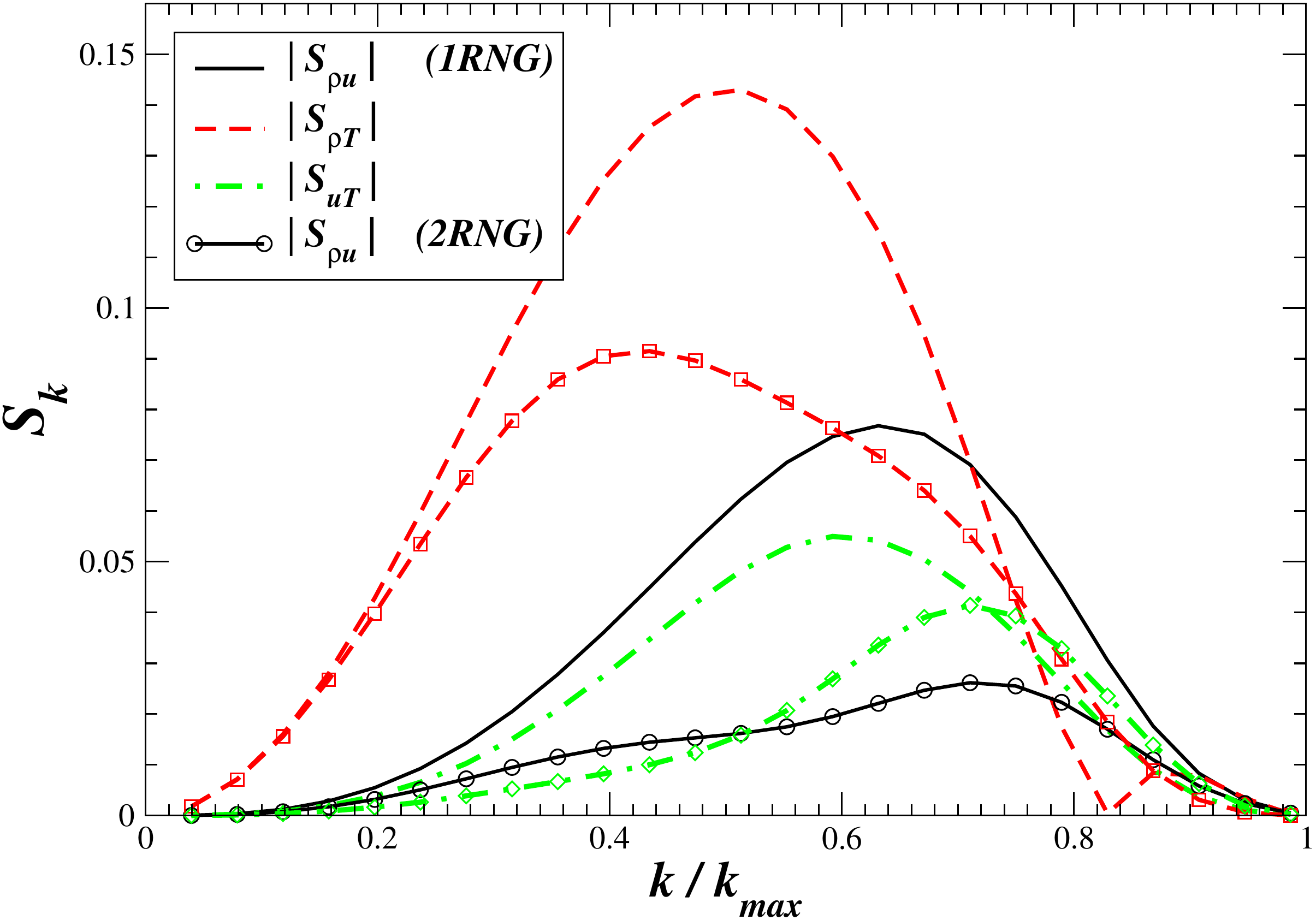}
\par\end{centering}

\caption{\label{fig:S_k_RK3}An illustration of the discrete structure factor
$\V{S}_{k}$ for the LLNS equation for the RK3-1RNG (lines) and RK3-2RNG
(same style of lines with added symbols) schemes, as calculated by
numerical solution of (\ref{S_k_equation_general}) for an ideal one
dimensional gas, for $\alpha=0.5$, $\beta=0.2$ and $\beta_{T}=0.1$.
(Left) Diagonal (self) structure factors, which should ideally be
identically unity. Also shown is the leading order error term $1+\epsilon(\alpha)\D{k}^{2}$
(dotted line), which is the same for both schemes. (Right) Off-diagonal
(cross) structure factors, which should ideally be identically zero.}

\end{figure}

We can now theoretically study the behavior of the RK3-1RNG and RK3-2RNG
schemes on the full linearized system (\ref{LLNS_1D_ideal}), specializing
to the case of zero background flow, $v_{0}=0$. As expected, we find
that the behavior is very similar to the one observed for the advection-diffusion
equation, in particular, the leading order terms have the same basic
form. Specifically, the expansions of the diagonal and off-diagonal
components of the structure factor $\M{S}_{\ki}$ for the RK3-1RNG
method are\begin{align}
S_{\ki}^{(\rho)}\approx S_{\ki}^{(T)}\approx1+\frac{S_{\ki}^{(u)}-1}{3}\approx & 1+\epsilon(\alpha)\D{k}^{2}\nonumber \\
S_{\ki}^{(\rho,u)}\approx & \frac{i}{12r}\alpha^{2}\D{k}^{3}\nonumber \\
S_{\ki}^{(\rho,T)}\approx & 2\epsilon(\alpha)\D{k}^{2}\nonumber \\
S_{\ki}^{(u,T)}\approx & i\frac{r-p}{6pr}\alpha^{2}\D{k}^{3},\label{S_k_diff_adv_RK3}\end{align}
where \[
\epsilon(\alpha)=-\frac{3\alpha^{3}pr}{4\left(3p+2r\right)}.\]
These structure factors are shown in Fig. \ref{fig:S_k_RK3} for sample
discretization parameters, along with the corresponding results for
RK3-2RNG. We see from these expressions that as the speed of sound
dominates the stability restrictions on the timestep more and more,
namely, as $p$ or $r$ become larger and larger, a smaller $\alpha$
is required to reach the same level of accuracy, that is, a smaller
timestep relative to the acoustic CFL stability limit is required.

Similar results to Eqs. (\ref{S_k_diff_adv_RK3}) hold also for the
isothermal LLNS equations (in which the there is no energy equation),
for which the calculations are simpler. For linearization around a
constant background flow of speed $v_{0}=c_{0}\mbox{Ma}$, where $\mbox{Ma}$
is the reference Mach number, the analysis for the isothermal LLNS
equations shows that the error grows with the Mach number as\[
S_{\ki}^{(\rho)}\approx1+\epsilon(\alpha)\left[1+6\mbox{Ma}^{2}+\mbox{Ma}^{4}\right]\D{k}^{2}.\]

\section{\label{sec:Higher-Dimensions}Higher Dimensions}

Much of what we already described for one dimension applies directly
to higher dimensions \citet{Bell:07,Bell:09}. However, there is a
peculiarity with the LLNS equations in three dimensions that does
not appear in one dimension, and also does not appear for the scalar
diffusion equation \citet{AMR_ReactionDiffusion_Atzberger}. In one
dimension the velocity component of the LLNS system of equations is
essentially an advection-diffusion equation. In higher dimensions,
however, there is an important difference, namely, the dissipation
operator is a \emph{modified} Laplacian $\M{\mathcal{L}}_{m}$. By
neglecting the hyperbolic coupling between velocity and the other
variables in the linearized LLNS equations, we obtain the \emph{stochastic
diffusion equation}\begin{equation}
\V{\vartheta}_{t}=\eta\grad\cdot\left[\M{C}(\grad\V{\vartheta})\right]+\sqrt{2\eta}\grad\cdot\left[\M{C}^{1/2}\V{\mathcal{W}}\right]=\eta\left(\M{\mathcal{D}}\M{C}\M{\mathcal{G}}\right)\V{\vartheta}+\sqrt{2\eta}\M{\mathcal{D}}\M{C}^{1/2}\V{\mathcal{W}}=\eta\M{\mathcal{L}}_{m}\V{\vartheta}+\sqrt{2\eta}\V{\mathcal{W}}_{m},\label{stoch_diffusion_SPDE}\end{equation}
where $\M{C}$ is the linear operator that transforms the velocity
gradient into a traceless symmetric stress tensor,\[
\M{C}(\grad\V{\vartheta})=2\left[\frac{1}{2}(\grad\V{\vartheta}+\grad\V{\vartheta}^{T})-\frac{\M{I}}{3}\left(\grad\cdot\V{\vartheta}\right)\right],\]
and we have denoted the continuum velocity field with $\V{\vartheta}\equiv\M{\mathcal{U}}$
in order to distinguish from the discretized velocities $\V{v}\equiv\V{U}$.
Here we will focus on two-dimensional flows, $\V{\vartheta}=\left[\vartheta_{x},\vartheta_{y}\right]$,
however, identical considerations apply to the fully three-dimensional
case.

If we arrange the components of the velocity gradient as a vector
with four components\[
\grad\V{\vartheta}=\left[\begin{array}{cccc}
\partial_{x}\vartheta_{x}, & \partial_{x}\vartheta_{y}, & \partial_{y}\vartheta_{x}, & \partial_{y}\vartheta_{y}\end{array}\right]^{T},\]
the linear operator $\M{C}$ in (\ref{stoch_diffusion_SPDE}) becomes
the matrix\begin{equation}
\M{C}=\left[\begin{array}{cccc}
\frac{4}{3} & 0 & 0 & -\frac{2}{3}\\
0 & 1 & 1 & 0\\
0 & 1 & 1 & 0\\
-\frac{2}{3} & 0 & 0 & \frac{4}{3}\end{array}\right],\label{Covariance_quasi2D}\end{equation}
which is not diagonal. This means that the components of the stochastic
stress $\M{C}^{1/2}\V{W}$ would need to have non-trivial correlations
between the $x$ fluxes for $v_{x}$ and $y$ fluxes for $v_{y}$,
as well as between the $x$ fluxes for $v_{y}$ and $y$ fluxes for
$v_{x}$. These correlations essentially amount to the requirement
that the stochastic stress be a traceless symmetric tensor, at least
at the level of its covariance matrix. Numerically, one generates
independent random variates for the upper triangular portion of the
stochastic stress tensor for each cell, then makes the tensor traceless
and symmetric \citet{LLNS_Espanol}. Note that one can save one random
number by using only $d-1$ variates to generate the diagonal elements.

However, it is important to point out that an \emph{equivalent }formulation
is obtained by using the operator\begin{equation}
\M{C}=\left[\begin{array}{cccc}
\frac{4}{3} & 0 & 0 & \frac{1}{3}\\
0 & 1 & 0 & 0\\
0 & 0 & 1 & 0\\
\frac{1}{3} & 0 & 0 & \frac{4}{3}\end{array}\right]=\M{I}+\left[\begin{array}{cccc}
\frac{1}{3} & 0 & 0 & \frac{1}{3}\\
0 & 0 & 0 & 0\\
0 & 0 & 0 & 0\\
\frac{1}{3} & 0 & 0 & \frac{1}{3}\end{array}\right],\label{Covariance_split}\end{equation}
where there is non-trivial cross correlations only between the $x$
fluxes for $v_{x}$ and $y$ fluxes for $v_{y}$. The splitting of
the operator $\M{C}$ in (\ref{Covariance_split}) corresponds to
rewriting the the stochastic diffusion equation (\ref{stoch_diffusion_SPDE})
in the equivalent but suggestive form\begin{eqnarray}
\V{\vartheta}_{t} & = & \eta\left[\grad^{2}\V{\vartheta}+\frac{1}{3}\grad\left(\grad\cdot\V{\vartheta}\right)\right]+\sqrt{2\eta}\left[\left(\grad\cdot\V{\mathcal{W}}_{T}\right)+\sqrt{\frac{1}{3}}\grad\mathcal{W}_{V}\right]\nonumber \\
 & = & \eta\left(\M{\mathcal{D}}_{T}\M{\mathcal{G}}_{T}+\frac{1}{3}\M{\mathcal{G}}_{V}\M{\mathcal{D}}_{V}\right)\V{\vartheta}+\sqrt{2\eta}\left(\M{\mathcal{D}}_{T}\V{\mathcal{W}}_{T}+\sqrt{\frac{1}{3}}\M{\mathcal{G}}_{V}\mathcal{W}_{V}\right),\label{split_stoch_diffusion_SPDE}\end{eqnarray}
where we have now distinguished between the \emph{tensorial} divergence
$\M{\mathcal{D}}_{T}$ and gradient operators $\M{\mathcal{G}}_{T}=\M{\mathcal{D}}_{T}^{\star}$,
which map from tensor to vector fields and vector to tensor fields,
respectively, and the \emph{vectorial} divergence $\M{\mathcal{D}}_{V}$
and gradient operators $\M{\mathcal{G}}_{V}=\M{\mathcal{D}}_{V}^{\star}$,
which map from vector to scalar fields and scalar to vector fields,
respectively. Corresponding to the splitting of the modified Laplacian
$\M{\mathcal{L}}_{m}=\M{\mathcal{D}}\M{C}\M{\mathcal{G}}=\M{\mathcal{L}}_{T}+\M{\mathcal{L}}_{V}$
into the tensorial Laplacian operator $\M{\mathcal{L}}_{T}=\M{\mathcal{D}}_{T}\M{\mathcal{G}}_{T}$
and the vectorial component $\M{\mathcal{L}}_{V}=\M{\mathcal{G}}_{V}\M{\mathcal{D}}_{V}/3$,
in Eq. (\ref{split_stoch_diffusion_SPDE}) we have split the stochastic
stress into a tensor white-noise field $\V{\mathcal{W}}_{T}$ in which
all components are uncorrelated, and a scalar white-noise field $\mathcal{W}_{V}$,
which we will call the stochastic \emph{divergence stress}. This representation
is perhaps more physically-intuitive than the standard formulation
in which the stochastic stress has unexpected exact symmetry and is
exactly traceless. Note that in the more general case where the diffusion
coefficient is spatially dependent and there is nonzero bulk viscosity
$\eta_{B}$, the dissipative term in (\ref{split_stoch_diffusion_SPDE})
becomes $\grad\cdot\left[\eta(\grad\V{\vartheta})\right]+\grad\left[\left(\eta/3+\eta_{B}\right)\grad\cdot\V{\vartheta}\right]$,
with an equivalent change in the stochastic term. Also note that for
the fluctuating incompressible Navier-Stokes equation the term with
the velocity divergence disappears and the dissipation operator is
a projected traditional Laplacian \citet{StochasticImmersedBoundary,SELM},
while the stochastic flux is simply a projected tensor white-noise
field.

\subsection{Discrete Fluctuation Dissipation Balance}

Our ultimate goal is to find a scheme that satisfies the discrete
fluctuation dissipation theorem, that is, find a discrete modified
Laplacian $\M{L}_{m}$ that is a consistent approximation to the continuum
modified Laplacian $\M{\mathcal{L}}_{m}(\V{k})\widehat{\V{\vartheta}}=\V{k}\cdot\left[\M{C}(\V{k}\widehat{\V{\vartheta}}^{T})\right]$
for small $k$, and a way to efficiently generate random increments
$\M{W}_{m}$ that discretize $\V{\mathcal{W}}_{m}$ and whose covariance
is $\left\langle \M{W}_{m}\M{W}_{m}^{\star}\right\rangle =\M{L}_{m}$.
This task is non trivial in general, and completing it requires some
ingenuity and insight, as illustrated in the work of Atzberger on
multigrid methods for the scalar stochastic diffusion equation \citet{AMR_ReactionDiffusion_Atzberger}.
We illustrate two different approaches next, the first corresponding
to attempting to directly discretize the modified Laplacian $\M{\mathcal{L}}_{m}$,
and the second corresponding to discretizing the split Laplacian $\M{\mathcal{L}}_{T}+\M{\mathcal{L}}_{V}/3$.
In the continuum context these are, of course, equivalent, but this
is not the case in the discrete context. Namely, in the continuum
formulation, $\M{C}$ maps from gradients to stresses, the divergence
operator $\M{\mathcal{D}}$ maps from fluxes to fields, and the gradient
$\M{\mathcal{G}}$ maps from fields to gradients. In the continuum
context, stresses, gradients and fluxes are all tensor fields and
thus in the same Hilbert space. In the discrete context, however,
stresses, gradients and fluxes may be discretized differently and
thus belong to different spaces.

\subsubsection{\label{Section_L_m}The modified Laplacian approach}

One approach to the problem of constructing discrete operators that
satisfy the discrete fluctuation-dissipation balance is to find a
discretization of the divergence $\M{D}$ and gradient $\M{G}$ operators
that are skew-adjoint and then form the modified Laplacian $\M{L}_{m}=\M{D}\M{C}\M{G}=-\M{D}\M{C}\M{D}^{\star}$,
and generate the stochastic increments as $\V{W}_{m}=\M{D}\M{C}^{1/2}\V{W}$.
As discussed above, for the meaning of $\M{C}^{1/2}$ to be clear,
stresses and gradients must belong to the same space. Furthermore,
it is required that the discrete operators $\M{D}$ and $\M{G}$ be
skew adjoint so that the discrete fluctuation dissipation balance
condition (\ref{discrete_FD_balance}) is satisfied.

The issue of how to define skew adjoint $\M{D}$ and $\M{G}$ operators
also arose in the historical development of projection algorithms
for incompressible flow. The incompressible flow literature suggests
two approaches that discretize both gradients and stresses by representing
them with tensors at the same grid of points. The first approach corresponds
to fully cell-centered discretization originally proposed by Chorin
\citet{Chorin68}, which uses centered differences to define a skew-adjoint
gradient and divergence operators. The second approach corresponds
to a finite element-based discretization developed by Fortin \citet{Fortin72}
and later used in the projection algorithm of Bell \emph{et al.} \citet{bellColellaGlaz:1989}.

In the Fortin approach both stresses and gradients are represented
as $d\times d$ tensors at the corners of a regular grid, where $d$
is the spatial dimension. The divergence operator $\M{D}$ combines
the values of the stresses at the $2d$ corners of a cell to produce
a value at the center of the cell. The gradient $\M{G}=-\M{D}^{\star}$
combines the values of the fields at the centers of the $2d$ cells
that share a corner into a gradient at that corner. In this scheme,
the stochastic stresses also live at the corners of the grid. They
are generated to have the required covariance, for example, (\ref{Covariance_quasi2D}).
Unfortunately, the discrete Fortin Laplacian $\M{L}=\M{D}\M{G}$ suffers
from a serious drawback: It has a nontrivial null space. For example,
for the scalar heat equation on a uniform grid in two dimensions,
the Laplacian stencil obtained from the Fortin discretization is\[
\left(L^{(F)}u\right)_{i,j}=\D{x}^{-2}\left[\frac{1}{2}\left(u_{i+1,j+1}+u_{i-1,j+1}+u_{i-1,j-1}+u_{i+1,j-1}\right)-2u_{i,j}\right],\]
for which the odd ($i+j$ odd) and even ($i+j$ even) points on the
grid are completely decoupled. In Fourier space the above Laplacian
is $-2\left[1-\cos\left(\D{k}_{x}\right)\cos\left(\D{k}_{y}\right)\right]$
and thus vanishes for the largest wavevectors, $\left|\D{k}_{x}\right|=\pi$,
$\left|\D{k}_{y}\right|=\pi$, which correspond to checker board zero
eigenmodes.

It can easily be verified that the same type of checker board zero
eigenmodes also exist for the modified Fortin Laplacian $\M{L}_{m}=\M{D}\M{C}\M{G}$.
In three dimensions, there are $O(N)$ zero eigenmodes for a grid
of size $N^{3}$. Issues arising when using these types of stencils
in the deterministic context are discussed in Almgren \emph{et al}.
\citet{almgrenBellSzymczak:1996}. Our theory for the structure factor
implicitly relies on the definiteness of the discrete generator, and
in fact, in the general non-linear setting the zero modes lead to
instabilities of the solution of the full LLNS system of equations.
We therefore abandon the Fortin corner-centered discretization of
the fluxes. 

Fully cell-centered approximations to $\M{D}$ and $\M{G}$ based
on second-order centered differences, previously studied in the context
of projection methods for incompressible flows by Chorin \citet{Chorin68},
lead to a discrete Laplacian that also has a non-trivial null space
and suffers similar shortcomings as the Fortin Laplacian. Specifically,
even in one dimension one obtains a Laplacian stencil\[
\left(L^{(C)}u\right)_{i}=\frac{1}{4\D{x}^{2}}\left[u_{i-2}-2u_{i}+u_{i+2}\right]\]
where the odd-even decoupling is evident. Here we develop a cell-centered
(collocated) discretization that preserves the null space of the continuum
Laplacian.

\subsubsection{\label{Section_split_L}The split Laplacian approach}

An alternative to trying to form a discrete modified Laplacian $\M{L}_{m}=\M{L}_{T}+\M{L}_{V}$
directly is to use the splitting in Eq. (\ref{split_stoch_diffusion_SPDE})
and form the discrete tensorial $\M{L}_{T}=\M{D}_{T}\M{G}_{T}$ and
vectorial $\M{L}_{V}=\M{G}_{V}\M{D}_{V}/3$ components separately
from discretizations of the tensorial and vectorial divergence and
gradient operators that are skew-adjoint, $\M{G}_{T}=\M{D}_{T}^{\star}$
and $\M{G}_{V}=\M{D}_{V}^{\star}$. The stochastic increments would
simply be generated as $\M{D}_{T}\V{W}_{T}+\M{G}_{V}W_{V}/\sqrt{3}$,
where $W_{V}$ and the components of $\V{W}_{T}$ are independent
normal variates.

A popular approach to discretizing the tensorial divergence and gradient
operators, commonly referred to as a MAC discretization in projection
algorithms for incompressible flow \citet{HarWel65}, defines a divergence
at cells centers from normal fluxes on edges, with a corresponding
gradient that gives normal derivatives at cell edges from cell-centered
values\begin{align}
\left(\M{D}\M{Z}\right)_{i,j}= & \D{x}^{-1}\left(\V{Z}_{i+\frac{1}{2},j}^{(x)}-\V{Z}_{i-\frac{1}{2},j}^{(x)}\right)+\D{y}^{-1}\left(\V{Z}_{i,j+\frac{1}{2}}^{(y)}-\V{Z}_{i,j-\frac{1}{2}}^{(y)}\right)\rightarrow\grad\cdot\V{Z}\label{D_discretization_facial}\\
-\left(\M{D}^{\star}\M{v}\right)_{i+\frac{1}{2},j}= & \D{x}^{-1}\left(\V{v}_{i+1,j}-\V{v}_{i,j}\right)\rightarrow\frac{\partial\V{v}}{\partial x}\nonumber \\
-\left(\M{D}^{\star}\M{v}\right)_{i,j+\frac{1}{2}}= & \D{y}^{-1}\left(\V{v}_{i,j+1}-\V{v}_{i,j}\right)\rightarrow\frac{\partial\V{v}}{\partial y}.\nonumber \end{align}
In this discretization, the tensor field $\M{Z}=\left[\V{Z}^{(x)};\V{Z}^{(y)}\right]=\left[Z_{v_{x}}^{(x)},Z_{v_{y}}^{(x)};Z_{v_{x}}^{(y)},Z_{v_{y}}^{(y)}\right]$
is strictly divided into an $x$ vector $\V{Z}^{(x)}$, which is represented
on the $x$ faces of the grid, and a $y$ vector $\V{Z}^{(y)}$, represented
on the $y$ faces of the grid. The MAC discretization, which we used
in the earlier one-dimensional examples, leads to a standard 5 point
discrete Laplacian in 2D (3 point in 1D, 7 point in 3D),\[
\left(L^{(MAC)}u\right)_{i,j}=\left[\D{x}^{-2}\left(u_{i-1,j}-2u_{i,j}+u_{i+1,j}\right)+\D{y}^{-2}\left(u_{i,j-1}-2u_{i,j}+u_{i,j+1}\right)\right],\]
whose eigenvalue in Fourier space is $2\cos\left(\D{k}_{x}\right)+2\cos\left(\D{k}_{y}\right)-4$
and is strictly negative for all nonzero wavevectors, and thus does
not suffer from the instabilities of the Chorin and Fortin discrete
Laplacians, discussed in Section \ref{Section_L_m}.

The vectorial divergence and gradient operators cannot be discretized
using the MAC framework. Namely, $\M{D}_{V}$ must operate on a cell-centered
vector field $\M{v}$, whereas the MAC-type discretization operates
on face-centered values. Instead, for the vectorial component we can
use either the Chorin discretization \citet{Chorin68}, in which both
scalar and vector fields are both cell-centered, or the Fortin discretization
\citet{Fortin72}, in which scalar fields are represented at corners
and vector fields are cell-centered. Here we choose the Fortin discretization
and calculate a (scalar-valued) velocity divergence and the corresponding
divergence stress at the corners of the grid, and also generate a
(scalar) random divergence stress at each corner. The deterministic
and random components are added to form the total corner-centered
divergence stress, and the velocity increment is calculated from the
(vector-valued) cell-centered gradient of the divergence stresses.
Note that the nontrivial nullspace of $\M{L}_{V}$ does not pose a
problem since $\M{L}_{T}$ and thus also $\M{L}_{m}=\M{L}_{T}+\M{L}_{V}$
has a trivial nullspace.

The discrete modified Laplacian that is obtained by this mixed MAC/Fortin
discretization can be represented in terms of second-order centered-difference
stencils. The first (i.e., the $v_{x}$) component of this Laplacian
can be represented as a linear combination of the velocities in the
9 neighboring cells,\begin{align}
\left(\M{L}_{m}\V{v}\right)_{jk}^{(v_{x})}= & \sum_{l,m=-1}^{1}\left(\frac{1}{\D{x}^{2}}L_{2-m,2+l}^{(MAC,x)}v_{j+l,k+m}^{(x)}+\frac{1}{\D{y}^{2}}L_{2-m,2+l}^{(MAC,y)}v_{j+l,k+m}^{(x)}\right.\label{Laplacian_stencil_vx}\\
+ & \left.\frac{1}{3\D{x}^{2}}L_{2-m,2+l}^{(F,x)}v_{j+l,k+m}^{(x)}+\frac{1}{3\D{x}\D{y}}L_{2-m,2+l}^{(F,xy)}v_{j+l,k+m}^{(y)}\right),\nonumber \end{align}
where $\M{L}^{(MAC,x/y)}$ and $\M{L}^{(F,x/y)}$ correspond to a
second-order MAC and Fortin discretizations of the terms $\partial_{xx}\vartheta_{x}$
and $\partial_{yy}\vartheta_{y}$ respectively, and $\M{L}^{(F,xy)}$
discretizes $\partial_{xy}\vartheta_{y}$. The same stencils apply
to the second (i.e., the $v_{y}$) component of the Laplacian as well,
by symmetry,\begin{align}
\left(\M{L}_{m}\V{v}\right)_{jk}^{(v_{y})} & =\sum_{l,m=-1}^{1}\left(\frac{1}{\D{x}^{2}}L_{2-m,2+l}^{(MAC,x)}v_{j+l,k+m}^{(y)}+\frac{1}{\D{y}^{2}}L_{2-m,2+l}^{(MAC,y)}v_{j+l,k+m}^{(y)}\right.\label{Laplacian_stencil_vy}\\
+ & \left.\frac{1}{3\D{y}^{2}}L_{2-m,2+l}^{(F,y)}v_{j+m,k+l}^{(y)}+\frac{1}{3\D{x}\D{y}}L_{2-m,2+l}^{(F,xy)}v_{j+m,k+l}^{(x)}\right).\nonumber \end{align}
Note that we chose the peculiar indexing of the stencils so that when
printed on paper they correspond to the usual Cartesian representation
of the $xy$ grid. The coefficients of the MAC stencil (\ref{Laplacian_stencil_vx})
are\begin{equation}
\M{L}^{(MAC,x)}=\left[\begin{array}{ccc}
0 & 0 & 0\\
1 & -2 & 1\\
0 & 0 & 0\end{array}\right]\mbox{ and }\M{L}^{(MAC,y)}=\left[\begin{array}{ccc}
0 & 1 & 0\\
0 & -2 & 0\\
0 & 1 & 0\end{array}\right],\label{MAC_Laplacian_stencils}\end{equation}
while the Fortin stencils are\begin{equation}
\M{L}^{(F,x)}=\left[\begin{array}{ccc}
\frac{1}{4} & -\frac{1}{2} & \frac{1}{4}\\
\frac{1}{2} & -1 & \frac{1}{2}\\
\frac{1}{4} & -\frac{1}{2} & \frac{1}{4}\end{array}\right],\quad\M{L}^{(F,y)}=\left[\begin{array}{ccc}
\frac{1}{4} & \frac{1}{2} & \frac{1}{4}\\
-\frac{1}{2} & -1 & -\frac{1}{2}\\
\frac{1}{4} & \frac{1}{2} & \frac{1}{4}\end{array}\right],\mbox{ and }\M{L}^{(F,xy)}=\left[\begin{array}{ccc}
-\frac{1}{4} & 0 & \frac{1}{4}\\
0 & 0 & 0\\
\frac{1}{4} & 0 & -\frac{1}{4}\end{array}\right].\label{Fortin_Laplacian_stencils}\end{equation}

\subsection{Results in Three Dimensions}

Our theoretical calculations have helped in formulating a complete
three-stage Runge-Kutta scheme for solving the full LLNS system in
one, two or three spatial dimensions. We have discussed how to generate
stochastic fluxes in each stage, including the required correlations
among the components of the stochastic stress, and have also discussed
how to relate the stochastic fluxes in each stage. Since theoretical
calculation of the three-dimensional structure factors is out of reach,
we present some numerical results for the RK3-2RNG method in three
dimensions with the mixed MAC/Fortin handling of the split Laplacian
as given in Eqs. (\ref{Laplacian_stencil_vx}) and (\ref{Laplacian_stencil_vy}),
hereafter termed the \emph{RK3D-2RNG algorithm}. We note in passing
that it is also possible to discretize the modified Laplacian (see
Section \ref{Section_L_m}) using a MAC-like discretization of the
viscous and stochastic stresses that avoids the use of the Fortin
corner-based discretization of the divergence stress. This saves one
random number per cell per stochastic flux, however, it requires the
use of a non-standard randomized cell-to-face projection (splitting)
of the stochastic stresses that complicates the analysis and handling
of physical boundaries and makes parallelization more difficult. We
therefore do not describe this approach here, and only note that it
produces very similar structure factors to those reported here.

We focus on the behavior of the scheme in global equilibrium with
periodic boundary conditions. We have implemented the full non-linear
fluxes as proposed in Refs. \citet{Bell:07,Bell:09}, using the interpolation
in Eq. (\ref{PPM_interpolation}) for the hyperbolic fluxes and simple
interpolation of the spatially-varying viscosity and thermal conductivity
in the handling of the viscous and stochastic fluxes. However, in
the tests reported here we have made the magnitude of the fluctuations
small compared to the means to ensure that the behavior is very similar
to the linearized LLNS equations. Including the full non-linear system
guarantees conservation and ensures that there are no non-linearly
unstable modes. More careful study of the proper handling of non-linearity
in the LLNS equations themselves and the associated numerical solvers
is deferred to future publications; here, we focus on verification
that the nonlinear scheme produces behavior consistent with the linearized
analysis. We note that we have implemented the new RK3D algorithm
also for the LLNS equations for a mixture of two ideal gases, closely
following the original scheme described in Ref. \citet{Bell:09}.
We find that the spatial discretization satisfies the discrete fluctuation-dissipation
balance even in the presence of concentration as an additional primitive
variable and that the RK3D-2RNG method performs very well with reasonably-large
time steps.

\subsubsection{Static Structure Factors}

\begin{figure}[tbph]
\begin{centering}
\includegraphics[width=0.45\textwidth]{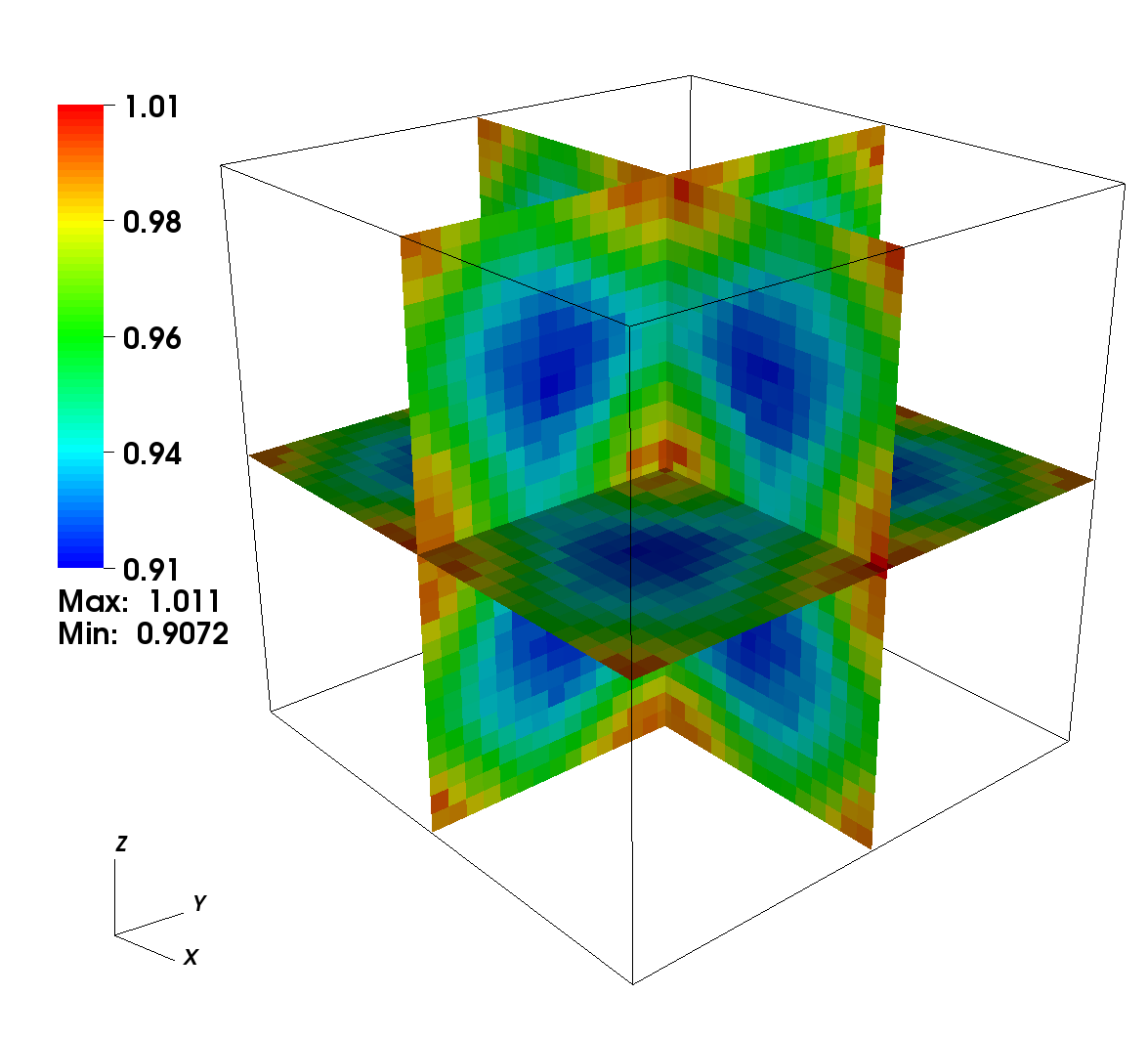}\includegraphics[width=0.45\textwidth]{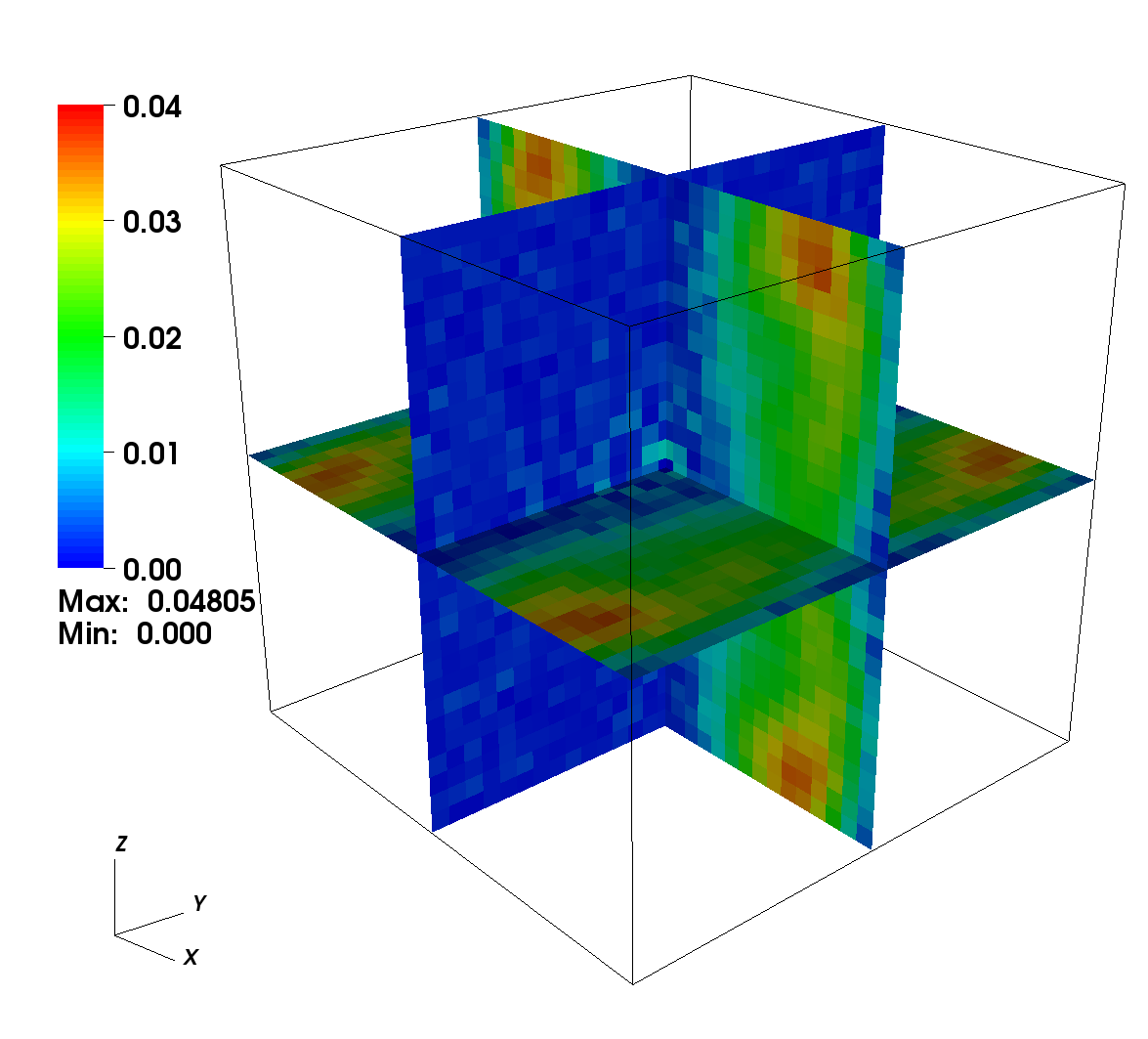}
\par\end{centering}

\begin{centering}
\includegraphics[width=0.45\textwidth]{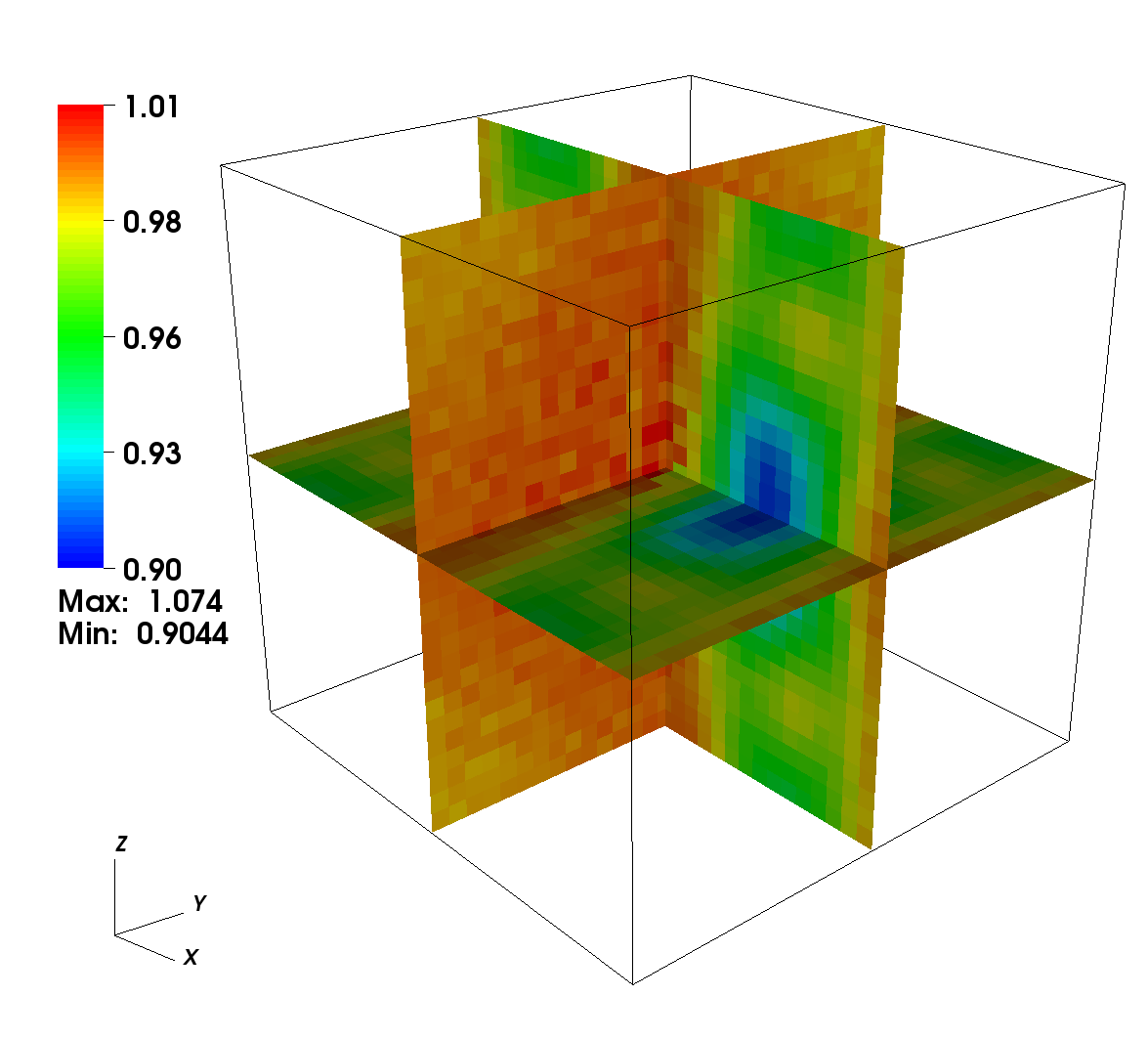}\includegraphics[width=0.45\textwidth]{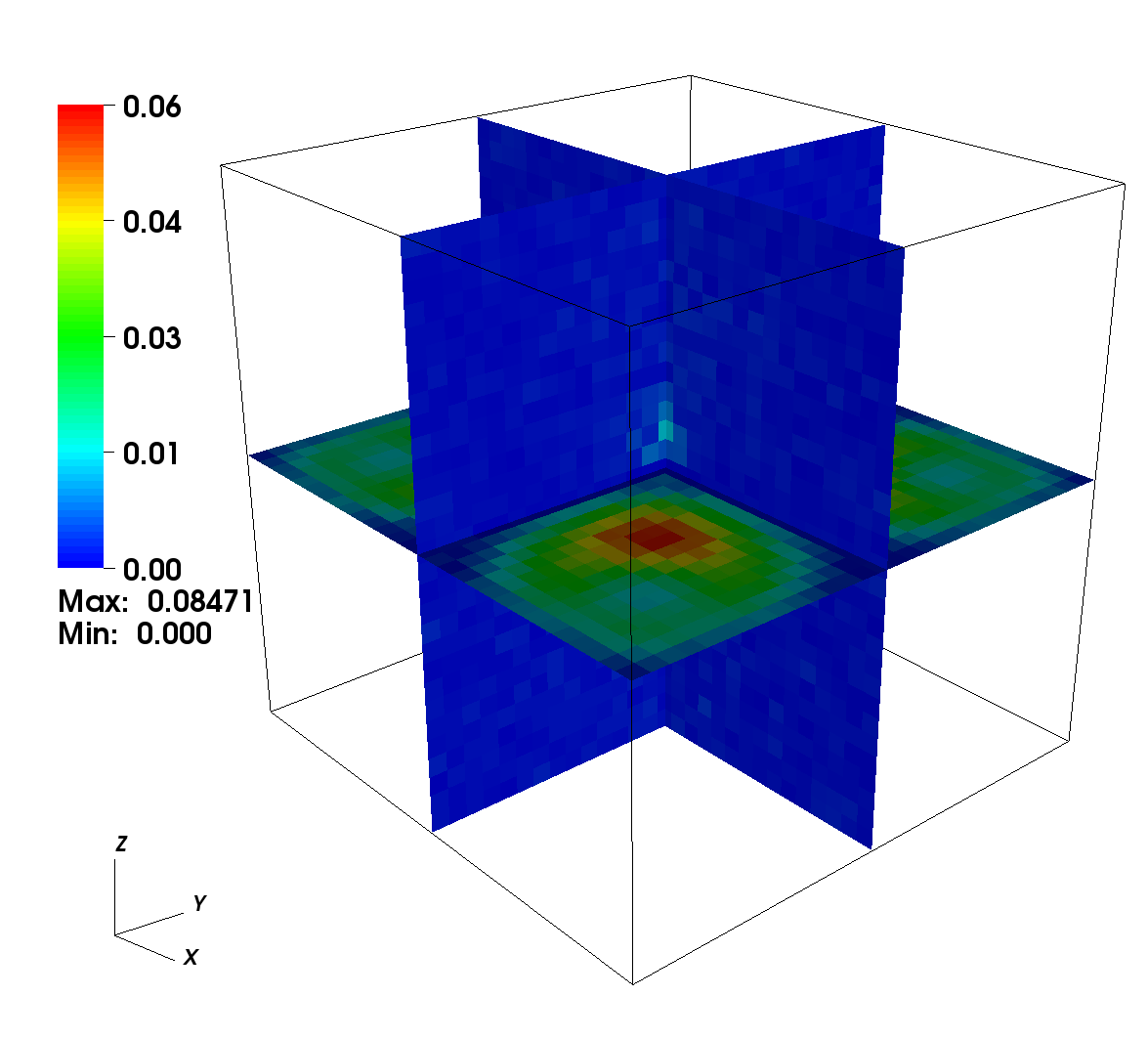}
\par\end{centering}

\begin{centering}
\includegraphics[width=0.45\textwidth]{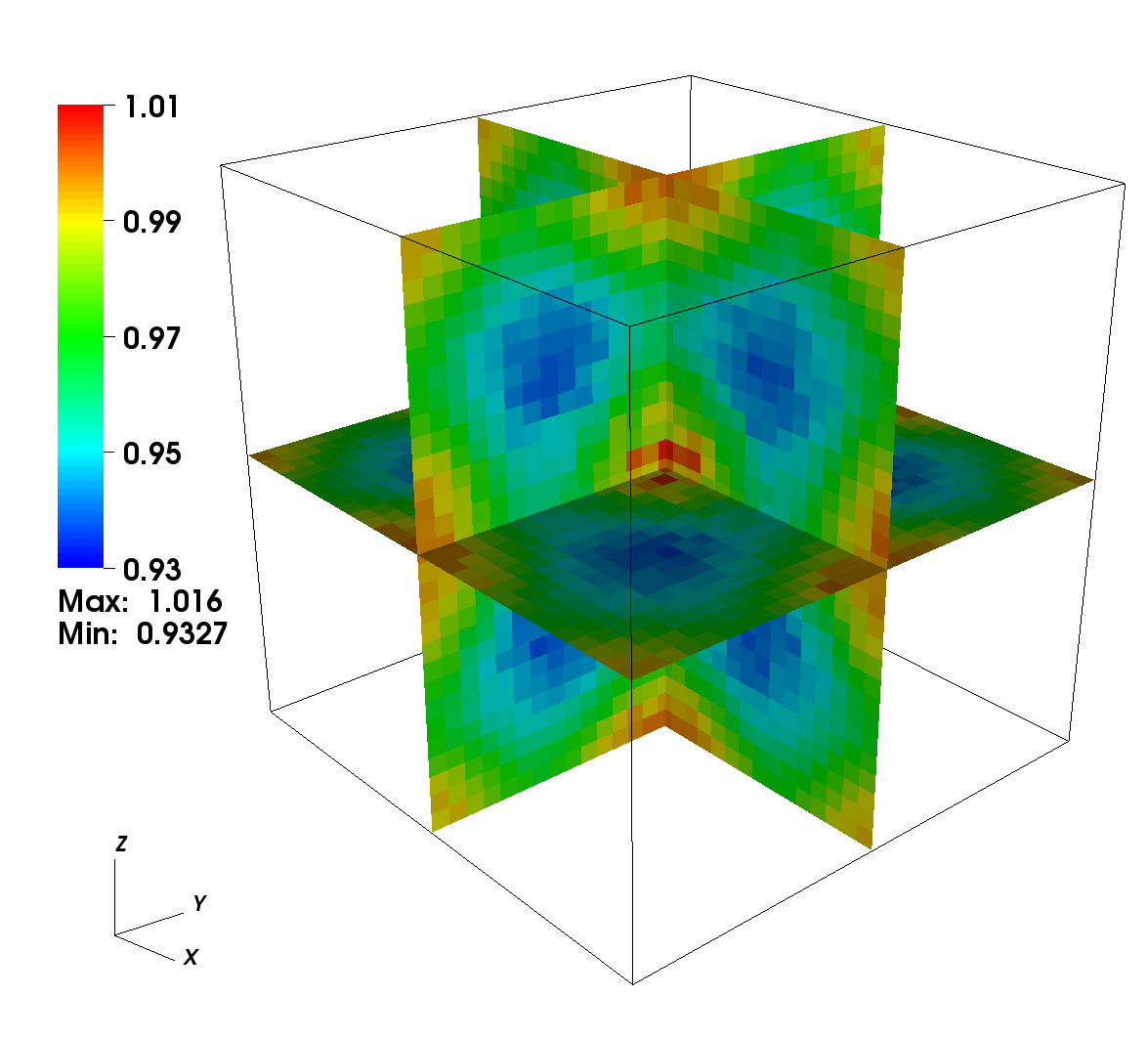}\includegraphics[width=0.45\textwidth]{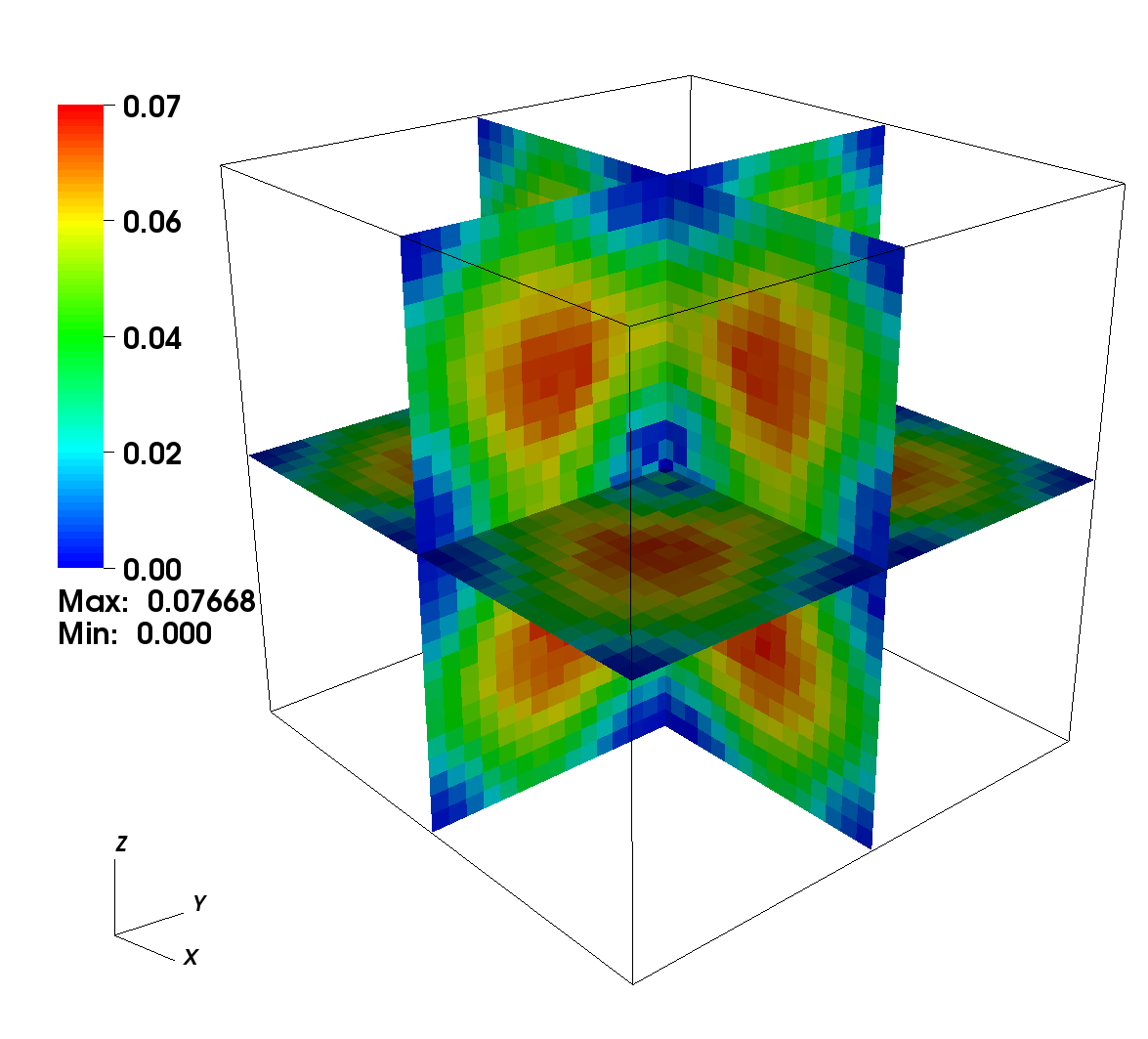}
\par\end{centering}

\caption{\label{RK3D.S_rho}(Left) $S_{\V{k}}^{(\rho)}$, $S_{\V{k}}^{(v_{x})}$
and $S_{\V{k}}^{(T)}$ (top to bottom); (Right) $\left|S_{\V{k}}^{(\rho,v_{x})}\right|$,
$\left|S_{\V{k}}^{(v_{x},v_{y})}\right|$ and $\left|S_{\V{k}}^{(\rho,T)}\right|$
(top to bottom) for RK3D-2RNG (Random Direction), with the time step
$\alpha=0.5$, $\beta=3\beta_{T}/2=0.1$, periodic boundary conditions
with $30^{3}$ cells, and averaging over $10^{6}$ time steps.}

\end{figure}

Examples of static structure factor $\M{S}_{\V{k}}$ for the RK3D-2RNG
scheme are shown in Fig. \ref{RK3D.S_rho}, showing that the diagonal
components $S_{\V{k}}^{(\rho)}$, $S_{\V{k}}^{(v_{x})}$ and $S_{\V{k}}^{(T)}$
are close to unity, while the off-diagonal components $S_{\V{k}}^{(\rho,v_{x})}$,
$S_{\V{k}}^{(v_{x},v_{y})}$ and $S_{\V{k}}^{(\rho,T)}$ are close
to zero (similar results hold for $S_{\V{k}}^{(v_{x},T)}$, not shown),
even for a large time step (half of the stability limit). Note that
the static structure factor is difficult to obtain accurately for
the smallest wavenumbers (slowest modes) and therefore the values
near the centers of the $\V{k}$-grid should be ignored.

It is seen in the figures that the diagonal components of $\M{S}_{\V{k}}$
are quite close to unity for the largest wavevectors, which is somewhat
surprising, and the largest error is actually seen for intermediate
wavenumbers, consistent with the one-dimensional results shown in
Fig. \ref{fig:S_k_RK3}. We have tested the method on several cell
Reynolds numbers $r$ and found that the results are worse as $r$
increases, consistent with the previous analysis, however, the higher
order of temporal accuracy allows for increasing the timestep to be
a reasonable fraction of the stability limit even for large $r$.

These results represent a significant improvement over the results
obtained for the original RK3 scheme presented in Bell \emph{et al.}
\citet{Bell:07,Bell:09}.\emph{ }Results with the original scheme
were sensitive to time steps, requiring small time steps to obtain
satisfactory results; the new scheme produces satisfactory results
for time steps near the stability limit. Also, through the use of
the mixed MAC and Fortin discretization, the new scheme eliminates
a weak but spurious correlation $S_{\V{k}}^{(v_{x},v_{y})}$ present
in the original scheme for small wavenumbers even in the limit of
small time steps.

\subsubsection{Dynamic Structure Factors}

\begin{figure}[tbph]
\begin{centering}
\includegraphics[width=0.45\textwidth]{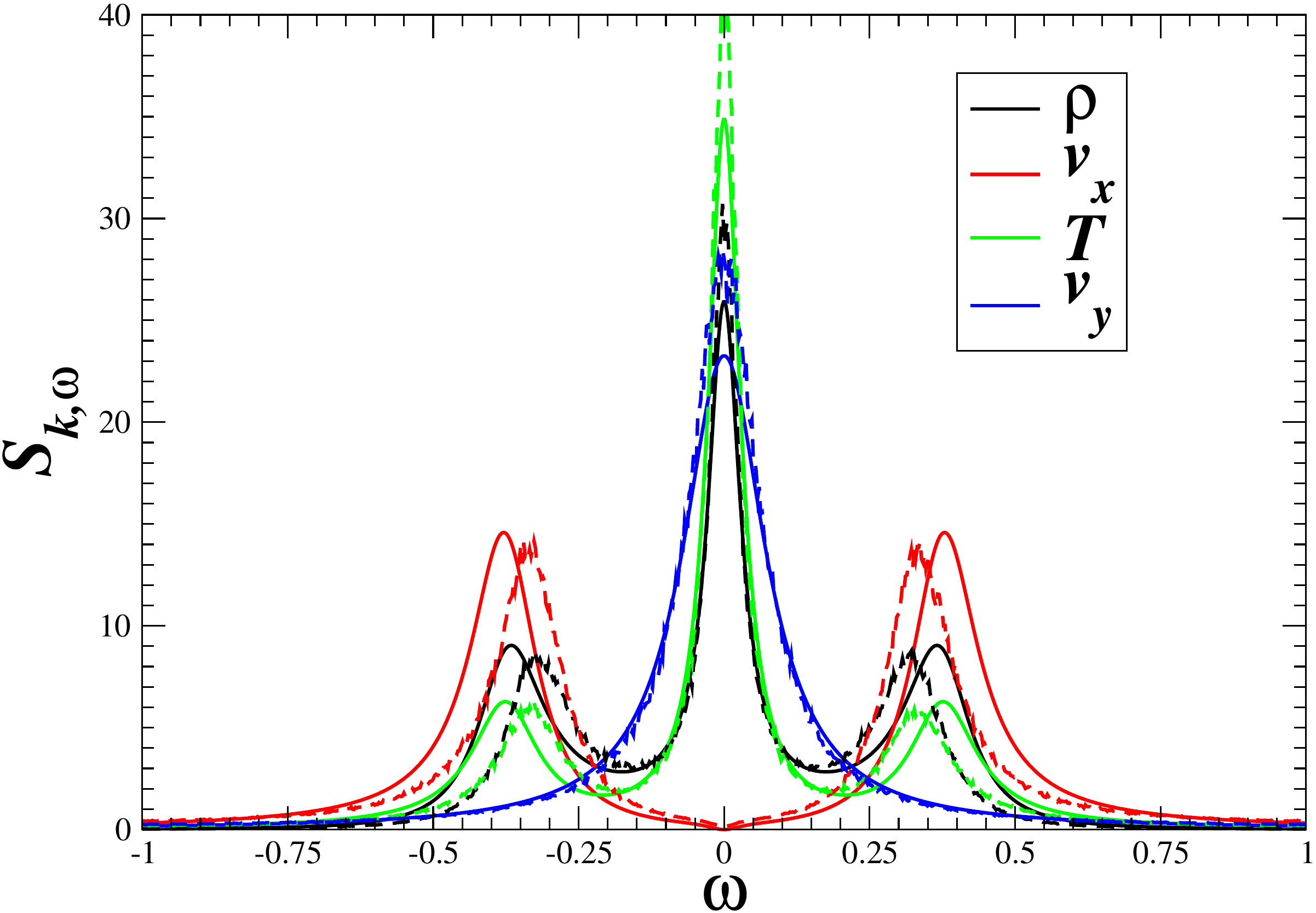}\includegraphics[width=0.45\textwidth]{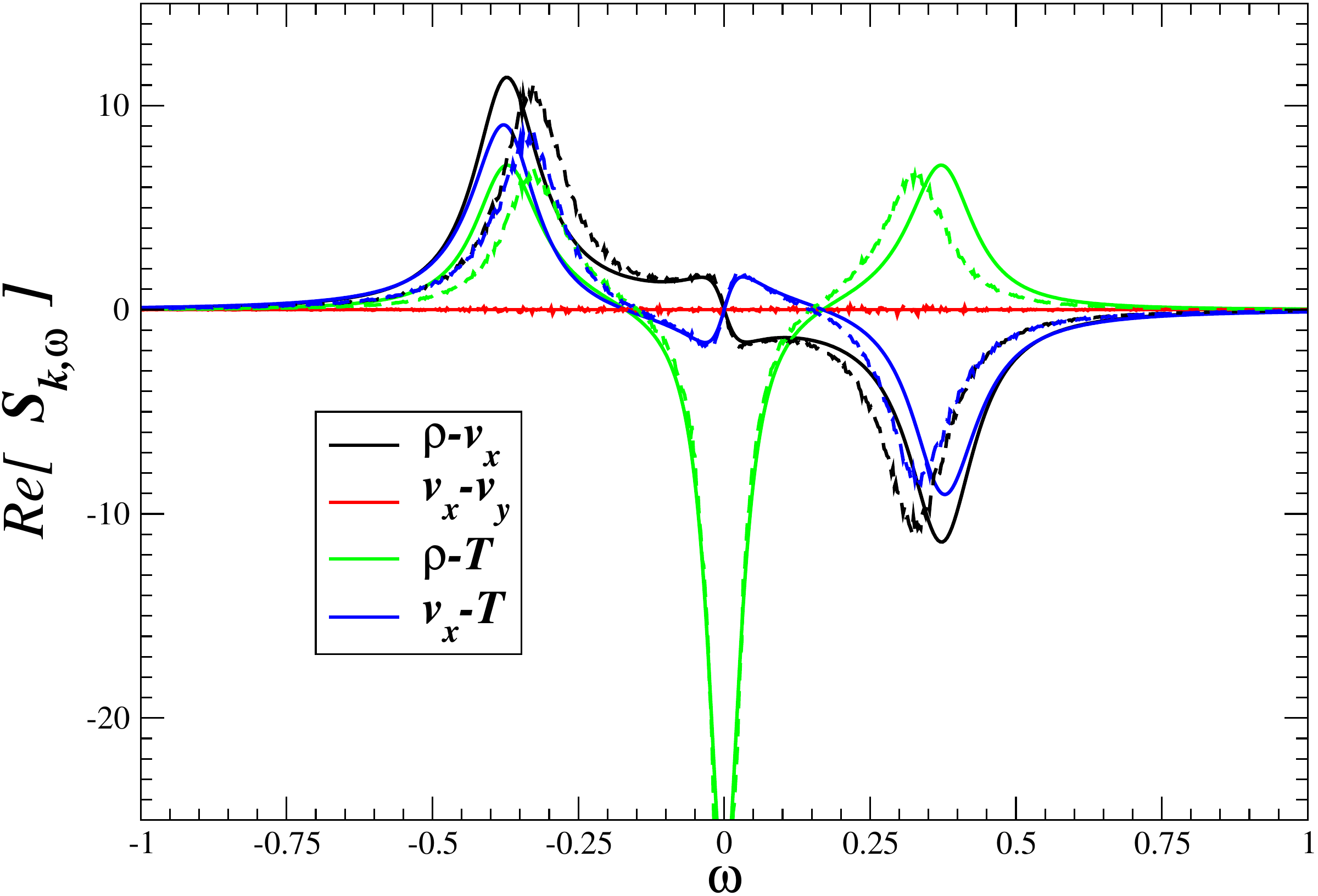}
\par\end{centering}

\begin{centering}
\includegraphics[width=0.45\textwidth]{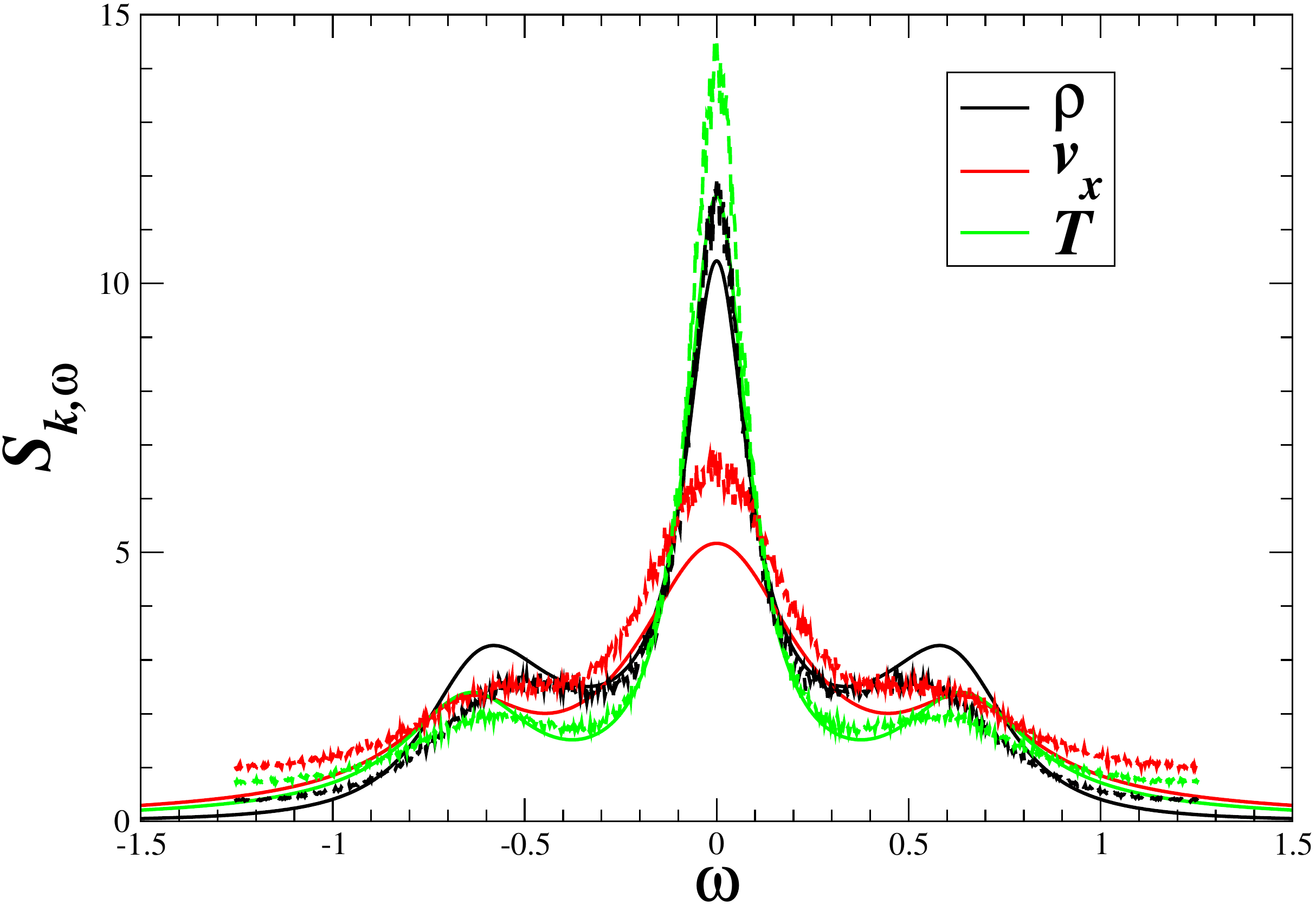}\includegraphics[width=0.45\textwidth]{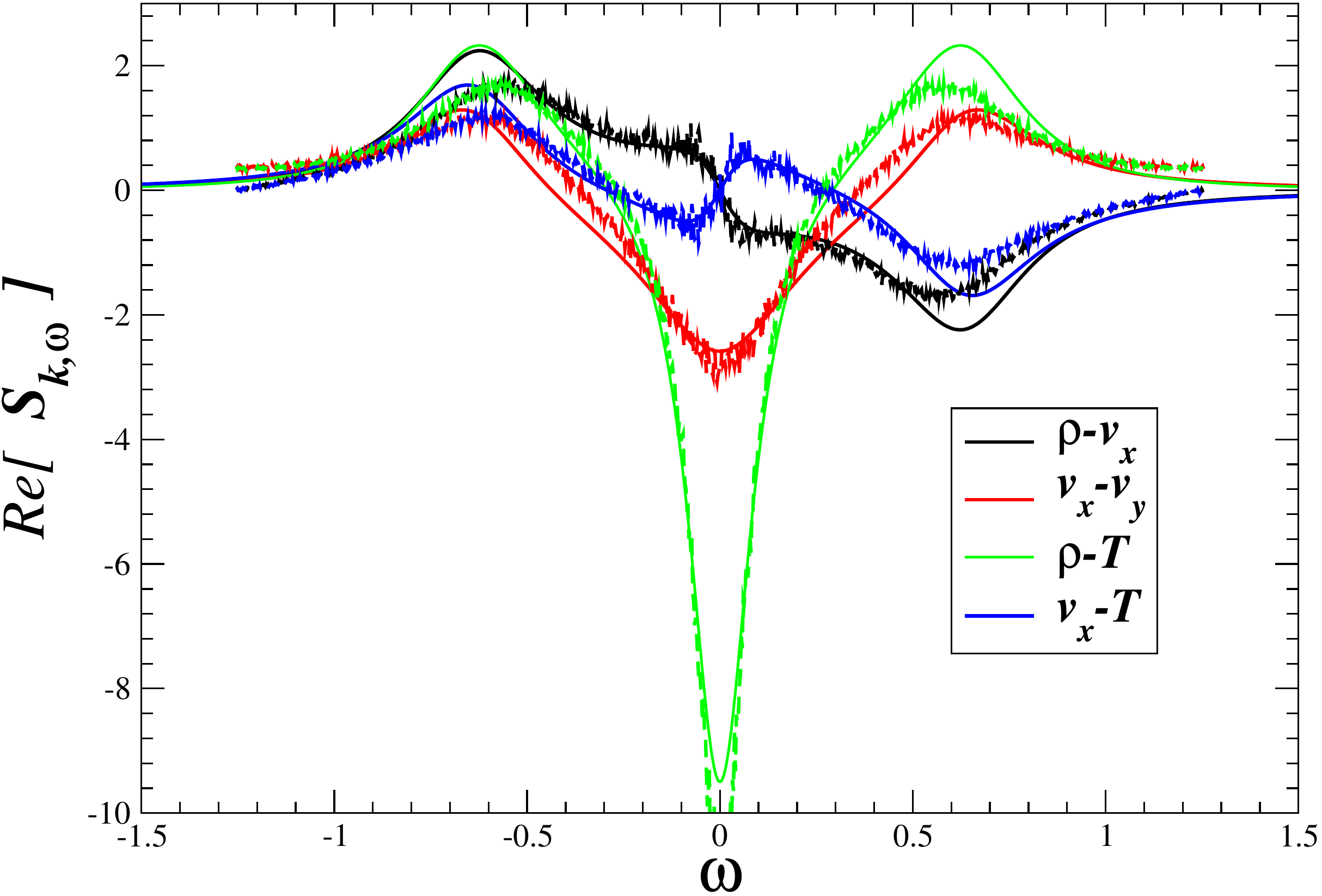}
\par\end{centering}

\caption{\label{RK3D.S_kw_diag_1}Diagonal (left) and the real part of the
off-diagonal (right) components of the dynamic structure factor $\M{S}_{\V{k},\omega}$
for RK3D-2RNG (dashed lines) for the same parameters as in Fig. \ref{RK3D.S_rho}.
For comparison, the analytical solution of the LLNS equations in Fourier
space are also shown (solid lines). The imaginary component of the
off-diagonal components is less than $0.1$ and it vanishes in the
theory. The top part shows the wavevector $\V{k=}\left(k_{max}/2,0,0\right)$
and the bottom shows the wavevector $\V{k=}\left(k_{max}/2,k_{max}/2,k_{max}/2\right)$.}

\end{figure}

Examples of dynamic structure factors $\M{S}_{\V{k},\wi}$ for the
RK3D-2RNG scheme are shown in Fig. \ref{RK3D.S_kw_diag_1} as a function
of $\omega$ for two relatively large wavevectors, along with the
correct continuum result obtained by solving the system (\ref{LLNS_linear_ideal})
through a space-time Fourier transform (we did not make any of the
usual approximations made in analytical calculations of $\M{S}_{\V{k},\wi}$
\citet{Zarate:07}, and instead used Maple's numerical linear algebra).
It is well-known that $\M{S}_{\V{k},\wi}^{(\rho)}$ and $\M{S}_{\V{k},\wi}^{(T)}$
exhibit three peaks for a given $\V{k}$ \citet{Zarate:07}, one central
Rayleigh peak at $\omega=0$ similar to the peak for the diffusion
equation {[}c.f. Eq. (\ref{theoretical_S_kw_diffusion})], and two
symmetric Brilloin peaks at $\omega\approx c_{s}k$, where $c_{s}$
is the adiabatic speed of sound, $c_{s}=c_{T}\sqrt{1+2/d_{f}}$ for
an ideal gas. For the velocity components, the transverse components
$\M{S}_{\V{k},\wi}^{(\V{v}_{\perp})}$ exhibit all three peaks, while
the longitudinal component $\M{S}_{\V{k},\wi}^{(v_{\parallel})}$
lacks the central peak, as seen in the figure. Note that as the fluid
becomes less compressible (i.e., the speed of sound increases), there
is an increasing separation of time-scales between the side and central
spectral peaks, showing the familiar numerical stiffness of the full
compressible Navier-Stokes equations.

We have verified that for small wavevectors the numerical dynamic
structure factors are in excellent agreement with the analytical predictions,
even for such large time steps. For wavevectors that are not small
compared to the discretization limits we do not expect a perfect dynamic
structure factor, even for very small time steps. It is important,
however, that the discretization behave reasonably for all wavevectors
(e.g., there should be no spurious maxima), and be somewhat accurate
for intermediate wavevectors, even for large time steps. As seen in
Fig. \ref{RK3D.S_kw_diag_1}, the RK3D-2RNG algorithm seems to perform
well even with a large time step. Improving the accuracy at larger
wavevectors requires using higher-order spatial differencing \citet{HighOrderFDSchemes}
(see discussion in Section \ref{sub:Higher-Order-Differencing}),
compact stencils (linear solvers) \citet{CompactFDStencils}, or pseudo-spectral
methods \citet{PseudoSpectralBook}, each of which has certain advantages
but also significant disadvantages over the finite-volume approach
in a more general nonlinear non-equilibrium context.

\section{Summary and Concluding Remarks}

In this paper we analyze finite volume schemes for the linearized
Landau-Lifshitz Navier-Stokes (LLNS) system (\ref{LLNS_linear_ideal})
and related SPDEs such as the stochastic advection-diffusion equation
(\ref{stoch_adv_diff_SPDE}). Our approach to studying the accuracy
of these explicit schemes is based on evaluating the discrete static
and dynamic structure factors, focusing on the accuracy at small wavenumber
$\D{k}=k\D{x}$. The methodology for formulating the structure factor
for numerical schemes is developed in sections \ref{sec:Explicit-Methods},
and then specialized to stochastic conservation laws in \ref{SectionLinearGeneral}.
Applying this analysis to the stochastic heat equation (\ref{stochastic_diffusion_SPDE})
in section \ref{sec:Section-Example:-Stochastic-Heat} we find the
truncation error for the Euler method to be $O(\D{t}k^{2})$; the
error for a standard predictor-corrector scheme is $O(\D{t}^{2}k^{4})$
using the same random numbers in the predictor and corrector stages
but $O(\D{t}^{3}k^{6})$ using independent random numbers at each
stage. Section \ref{sec:Section-LLNS-Equations-1D} extends this analysis
to the third-order Runge-Kutta scheme of Bell \emph{et al.} \citet{Bell:07,Bell:09}
for the one-dimensional advection-diffusion SPDE. We find the best
accuracy when the stochastic fluxes at the three stages are generated
from two sets of random numbers, as given by (\ref{RK3_optimal});
using this version, called RK3-2RNG, for the LLNS equations gives
good results, even when nonlinear effects are included (see figures
\ref{fig:S_k_RK3}, \ref{RK3D.S_rho}, and \ref{RK3D.S_kw_diag_1}).
Finally, section \ref{sec:Higher-Dimensions} explains why the cross-correlations
in the stress tensor in the three-dimensional LLNS require special
treatment and proposes a mixed MAC/Fortin discretization as a way
to obtain the desired discrete fluctuation-dissipation balance.

Here we have investigated linearized PDEs with stochastic fluxes where
the noise is additive. As such, the stability properties of the numerical
schemes are the same as for the deterministic case. Yet in practice
one would like to implement these schemes for the nonlinear stochastic
PDEs with state-dependent stochastic fluxes. While in the limit of
small fluctuations the behavior of the schemes is expected to be similar
to the linearized case, the proper mathematical foundation and even
formulation of the nonlinear fluctuating equations has yet to be laid
out. Furthermore, the stability properties of numerical schemes for
the nonlinear LLNS system are not well understood and the whole notion
of stability is different than it is for deterministic schemes. For
example, even at equilibrium, a rare fluctuation can cause a thermodynamic
instability (e.g., a negative temperature which implies a complex
sound speed) or a mechanical instability (e.g., a negative mass density).
Capping the noises in the stochastic flux terms will not necessarily
solve the problem because the hydrodynamic variables are time-correlated
so the numerical instability may not appear on a single step but rather
as an accumulated effect. We are investigating these issues and will
discuss strategies to address this type of stability issue in future
publications.

One of the advantages of finite volume solvers over spectral methods
is the ability to implement realistic, complex geometries for fluid
simulations. In this paper we only consider periodic boundaries but
many other boundary conditions are of interest, notably, impenetrable
flat hard walls with stick and slip conditions for the velocities
and either adiabatic (zero temperature gradient) or thermal (constant
temperature) conditions for the temperature. Equilibrium statistical
mechanics requires that the static structure factor be oblivious to
the presence of walls, even though the dynamic structure factors typically
exhibit additional peaks due to the reflections of fluctuations from
the boundaries. Therefore, the numerical discretization of the Laplacian
operator $\M{L}$, the divergence operator $\M{D}$ and the covariance
of the stochastic fluxes $\M{C}$ should continue to satisfy the discrete
fluctuation-dissipation balance condition $\M{L}+\M{L}^{\star}=-2\M{D}\M{C}\M{D}^{\star}$
and be consistent, even in the presence of boundaries. Standard treatments
of boundary conditions used in deterministic schemes can easily be
implemented in the stochastic setting \citet{Bell:07,AMR_ReactionDiffusion_Atzberger},
however, satisfying the discrete fluctuation-dissipation balance is
not trivial and requires modifying the stochastic fluxes and possibly
also the finite-difference stencils near the boundaries \citet{AMR_ReactionDiffusion_Atzberger}.
In particular, the case of Dirichlet boundary conditions is more complicated,
especially in the case of the mixed MAC and Fortin discretization
of the compressible Navier-Stokes equations. Complex boundaries present
further challenges even in the deterministic setting. We will explore
the issues associated will fluctuations at physical boundaries in
future publications.

One motivation for the development of numerical methods for the LLNS
equations is for their use in multi-algorithm hybrids. One emerging
paradigm in the modeling and simulation of multiscale problems is
Multi-Algorithm Refinement (MAR). MAR is a general simulation approach
that combines two or more algorithms, each of which is appropriate
for a different scale regime. MAR schemes typically couple structurally
different computational schemes such as particle-based molecular simulations
with continuum partial differential equation (PDE) solvers. The general
idea is to perform detailed calculations using an accurate but expensive
algorithm in a small region (or for a short time), and couple this
computation to a simpler, less expensive method applied to the rest.
The major difficulty is in constructing hybrid is that particle and
continuum methods treat noise in completely different ways. The challenge
is to ensure that the numerical coupling of the particle and continuum
computations is self-consistent, stable, and most importantly, does
not adversely impact the underlying physics. These problems become
particularly acute when one wants to accurately capture the physical
fluctuations at micro and mesoscopic scales. The correct treatment
of boundary conditions in stochastic PDE schemes is particularly difficult
yet crucial in hybrid schemes since the coupling of the two algorithms
is essentially a dynamic, two-way boundary condition. Recent work
by Tysanner \emph{et al.} \citet{EquilibriumReservoirs_Garcia}, Foo
\emph{et al.} \citet{Bell:06}, Williams \emph{et al. \citet{FluctuatingHydro_AMAR}}
and Donev \textit{et al.} \citet{DSMC_Hybrid} have demonstrated the
need to model fluctuations at the continuum level in hybrid continuum
/ particle approaches, however, a seamless coupling has yet to be
developed.

In this paper we consider the fully compressible LLNS system, for
many of the phenomena of interest the fluid flow aspects occur at
very low Mach numbers. Another topic of future work for stochastic
PDE schemes is to construct a low Mach number fluctuating hydrodynamics
algorithm. A number of researchers have considered extended versions
of the incompressible Navier Stokes equations that include a stochastic
stress tensor \citet{Moseler:00,Sharma:04,StochasticImmersedBoundary}.
This type of model does introduce fluctuations into the Navier Stokes
equations and is applicable in some settings, such as in modeling
simple Brownian motion. However, as pointed out by Zaitsev and Shliomis
\citet{Zaitsev:71}, the incompressible approximation introduces fictitious
correlations between the velocity components of the fluid. Furthermore,
this type of approach does not capture the full range of fluctuations
in the compressible equations. In particular, adding a stochastic
stress into the incompressible Navier Stokes equations creates fluctuations
in velocity but does not reproduce the large scale and slow fluctuations
in density and temperature, which persist even in the incompressible
limit. We plan to investigate alternative formulations that can capture
more of the features of the fluctuating hydrodynamics while still
exploiting the separation of scales inherent in low Mach number flows.
We also note that although the theoretical importance of distinguishing
between the incompressible approximation and the low-Mach number limit
is well-established for fluctuating hydrodynamics \citet{Bena:00,Zwanzig:75},
numerical algorithms for the latter have yet to be developed.

\begin{acknowledgments}
The authors wish to thank Berni Alder and Jonathan Goodman for helpful
discussions, and Paul Atzberger for inspiring perspectives on the
discrete fluctuation dissipation relation and a critical reading of
this paper. A. Donev's work was performed under the auspices of the
U.S. Department of Energy by Lawrence Livermore National Laboratory
under Contract DE-AC52-07NA27344. The work of J. Bell and A. Garcia
was supported by the Applied Mathematics Research Program of the U.S.
Department of Energy under Contract No. DE-AC02-05CH11231. The work
of E. Vanden-Eijnden was supported by the National Science Foundation
through grants NSF: DMS02-09959, DMS02-39625, and DMS07-08140, as
well as the Office of Naval Research through grant N00014-04-1-0565.
\end{acknowledgments}
\begin{appendix}

\subsection{\label{SectionImplicit}Semi-Implicit Crank-Nicolson Method}

When sound is included in the fluctuating hydrodynamic equations implicit
methods are not really beneficial since the large sound speed limits
the time step. However, for the pure stochastic diffusion/heat equation
or advection-diffusion equations with a small advection speed the
time step may become strongly limited by the diffusive CFL limit,
especially for small cells. In such cases an implicit method can be
used to lift the diffusive stability restriction on the time step.
For example, the second-order (in both space and time) Crank-Nicolson
semi-implicit scheme for the stochastic heat equation entails solving
the linear system \begin{align}
u_{j}^{n+1}-\frac{\mu\D{t}}{2\D{x}^{2}}\left(u_{j-1}^{n+1}-2u_{j}^{n+1}+u_{j+1}^{n+1}\right) & =\nonumber \\
u_{j}^{n}+\frac{\mu\D{t}}{2\D{x}^{2}}\left(u_{j-1}^{n}-2u_{j}^{n}+u_{j+1}^{n}\right) & +\sqrt{2\mu}\frac{\D{t}^{1/2}}{\D{x}^{3/2}}\left(W_{j+\frac{1}{2}}^{n}-W_{j-\frac{1}{2}}^{n}\right),\label{CrankNicolson_iteration}\end{align}
which is tridiagonal except at periodic boundaries.

The analysis carried out above for explicit schemes can easily be
extended to implicit methods since in Fourier space different wavevectors
again decouple and the above iteration becomes a scalar linear equation
for $\hat{u}_{k}^{n+1}$ that can trivially be solved. Firstly, it
is observed that the small time step limit is the same regardless
of the semi-implicit treatment, specifically, the same discrete fluctuation-dissipation
condition (\ref{discrete_FD_balance}) applies. Remarkably, for the
Crank-Nicolson iteration (\ref{CrankNicolson_iteration}) it is found
that the discrete static structure factor is independent of the time
step, $S_{\ki}=1$ for all $\beta$. The dynamic structure factor,
however, has the same spatial discretization errors (\ref{S_kw_zero_dt})
as for the Euler scheme even in the limit $\beta\rightarrow0$. Furthermore,
as expected, the dynamics is not accurate for large $\beta$ and the
time step cannot be enlarged much beyond the diffusive stability limit
related to the smallest length-scale at which one wishes to correctly
resolve the dynamics of the fluctuations.

If advection is included as well also discretized semi-implicitly,
the method again gives perfect structure factors, $S_{k}=1$ identically,
and is unconditionally stable. If only diffusion is handled semi-implicitly
but advection is handled with a predictor-corrector approach, then
it turns out that the optimal method is to not include a stochastic
flux in the predictor step, giving the same leading-order error term
as PC-2RNG in Eq. (\ref{PC_advection_results}) when $\left|r\right|>0$,
but giving a perfect $S_{k}=1$ when $r=0$.

\end{appendix}

%\bibliographystyle{unsrt}
%\bibliography{13_home_adonev_Papers_LLNS_References,14_home_adonev_Papers_LLNS_GarciaGeneralBibFile,15_home_adonev_Papers_LLNS_MScaleProp}

\end{document}